\documentclass{article}
\usepackage{multirow}
\usepackage{lineno}
\usepackage{float}
\modulolinenumbers[5]
\usepackage[utf8]{inputenc}
\usepackage[T1]{fontenc}
\usepackage[width=16cm,height=25cm]{geometry}

\usepackage{subfig,xcolor,graphicx}
\usepackage{amsmath,amssymb,amsfonts,amsthm}
\usepackage{newtxtext,newtxmath} 
\usepackage{amssymb}
\usepackage[algo2e,ruled,lined,titlenumbered,linesnumbered, algonl,boxed]{algorithm2e}
\SetCommentSty{normal}
\usepackage{hyperref}
\usepackage{hypcap}

\renewcommand{\vec}[1]{#1} %{\underline{#Fne1}}
\newcommand{\mat}[1]{#1} %{\underline{\underline{#1}}}
\title{Space/time  global/local noninvasive coupling strategy: application to viscoplastic structures}
  \author{Maxime Blanchard$^{a,b}$, Olivier Allix$^a$, Pierre Gosselet$^a$, Geoffrey Desmeure$^b$\\(a) 
LMT, ENS Paris-Saclay, CNRS, Universit\'{e} Paris-Saclay \\  (b) Safran Aircraft Engines }

\theoremstyle{remark}
\newtheorem{remark}{Remark}
\begin{document}

\maketitle

\begin{abstract}

The purpose of this paper is to extend the non-invasive global/local iterative coupling technique \cite{GEN09a} to the case of large structures undergoing nonlinear time-dependent evolutions at all scales. It appears that, due to the use of legacy codes, the use of different time grids at the global and local levels is mandatory in order to reach a satisfying level of precision. In this paper two strategies are proposed and compared for elastoviscoplastic models. 

The questions of the precision and performance of those schemes with respect to a monolithic approach is addressed. The methods are first exposed on a 2D example and then applied on a 3D part of industrial complexity.

\end{abstract}


\noindent  {This work was supported by Safran Aircraft Engines and is part of the MAIA-MM1 research program.}

\begin{center}\textbf{This work was published in Finite Elements in Analysis and Design, volume 156 pages 1-12 2019\\ doi:10.1016/j.finel.2019.01.003} 
\end{center}

\section*{Introduction}

The simulation of large structures where complex nonlinear local phenomena may occur is still a scientific and an industrial challenge. One of the main difficulties comes from the difference of length scales between the global response of the structure and the localized phenomena. 

To deal with it in an effective manner, concurrent multiscale methods have been developed. They are often based on domain decomposition techniques like, among others, the primal BDD method \cite{MAN93}, the dual FETI method \cite{FAR91} or the mixed Latin method \cite{LAD99}.

These methods are not currently available in legacy codes, even though different attempts have been made. This is due  both to the heavy development associated with their implementation and to their lack of robustness with respect to the variety of situations legacy codes have to deal with. 

When the small wavelength phenomena are localized in space, i.e. concerning a relatively reduced part of the whole body, multiscale computations may also be based on coupling techniques as for example the Arlequin method \cite{DHI98}. The implementation of such methods in a legacy code is also not straightforward, mainly because the creation of the coupling operators between the two models in the transition zone requires complex integration operations.

A method often used by engineers to tackle such problems is the submodeling approach \cite{COR99}: after a \emph{global} computation, structural zooms are applied on the \emph{local} critical zones, with details represented exactly. An advantage of this approach is that it allows to easily connect research software and commercial code, as was done for example to deal with the prediction of delamination under low velocity impacts \cite{allix_2001}.  Unfortunately, it implies neglecting the influence of the local zones on the whole structure. This may lead in turn to quite important local errors. The problem  becomes more crucial  when plasticity initiated locally spreads over the whole structure. 

\medskip

To correct the drawback of submodeling while keeping its simplicity and flexibility, a non-invasive method was proposed in order to allow exact \emph{local/global} analysis embedding the same basic tools as those used in the submodeling inside an iterative procedure.

The method relies on interface coupling; its formulation and its numerical optimization have first been derived in the case of global linear models and local plasticity \cite{GEN09a,GEN11}. A number of other applications and extensions have been proposed: use of XFEM at the local scale \cite{PAS13}, the treatment of non-matching interfaces \cite{LIU14}, coupling between a global plate model and 3D parts for bolted assemblies \cite{GUG14}, geometrically non conforming coupling \cite{GUI18}, multiscale time and space computation in explicit dynamic \cite{BET14-1} with implementation in Abaqus Explicit for the analysis of delamination under impact \cite{BET17}, non-invasive domain decomposition approach \cite{DUV16}. Mesh refinement based on error estimation may also be cast in the proposed non-intrusive framework \cite{DUV18}.

Alternative proposals exist, based on volume coupling. The volume coupling is performed either by means of a noninvasive version of the Partition of Unity method \cite{PLE12,FIL18}, by using projection techniques between the local and global models  \cite{TEM11,HOL13} or by means of homogenization like techniques \cite{HUE16}.
\medskip

So far the proposed method was applied for global linear models with local nonlinearities and topological changes. Safran Aircraft Engines was interested in exploring its potential for the detailed analysis of complex structures undergoing viscoplastic strains that can spread all over the structure. This situation is nowadays encountered in the case of aircraft engines. Indeed engines have undergone important improvements of their performance associated with the large increase of the working temperature and  the use of optimized designs involving very thin parts. For example, micro-perforations were added in order to improve the cooling of hot parts (submitted to an air flux at $1500^\circ$C), like the combustion chambers and the high pressure turbine blades. 

{\color{black} Our \textit{reference} problem, see Section~\ref{sec:3D} for a full description, is an elastoviscoplastic turbine blade where two zones of interest need particular mesh refinement in order to correctly represent a complex geometry. Even if in that relatively simple case a monolithic model could be assembled and computed, our aim is to adapt to (or maybe extend) the industrial practice and only make use of a \emph{global} representation of the blade (with simplified geometry in the zones of interest) and \emph{local} refined representations of the zones of interest. In that case, even if they occupy less than 10\% of the volume of the blade, the \textit{local} meshes for the zones of interest have significantly more nodes than the \textit{global} mesh (precisely three times more). We thus wish to derive an iterative coupling algorithm which converges to the {reference} (monolithic) solution by alternating \emph{elastoviscoplastic} {global} and {local} computations.}

\medskip

The quality of the integration of viscoplastic models is very sensitive to the size of the time steps. When applying the non-intrusive framework with global spreading of the viscoplastic strains, several difficulties occurred using legacy codes. They appeared to be related to {\color{black} the management of the precision of the integration of internal variables on the different models. Indeed each model needed its own adapted time discretization, which is controlled by} some automatic procedures, called cutbacks \cite{BLA17-2}.

Cutbacks correspond to the subdiscretization of a given time increment when convergence difficulties {\color{black} or precision issues} are encountered. {As the Global/Local coupling relies on different models and independent computations, cutbacks are likely to generate different increment histories, \textit{i.e.} different time discretizations, for the \emph{global} and \emph{local} models.} {\color{black} It is thus of importance to control the precision of all computations and to provide sufficient synchronization of the adapted time grids in order to preserve the coherence of the coupling.}

%When dealing with this problem, it appeared that the results we obtained relying on Abaqus default cutback procedure were most often unsatisfactory regarding the precision of the solution compared to a monolithic reference. The reason was that making use of {\color{black} the default settings} for the cutbacks did not guarantee {\color{black}sufficiently well} the quality of the integration in time and that different levels of accuracy were obtained for the \emph{global} and \emph{local} models. %Obviously, a good accuracy would be obtainable with Abaqus cutback technique but the calibration procedure would be very difficult and time consuming because of its strong dependence with respect to the model.

 \medskip

Based on this experience, the motivation of this paper is to propose and study different coupling strategies allowing to preserve the accuracy of the global/local coupling, {\color{black} making use of Abaqus' most advanced precision controls}.

The paper is organized as follow. In Section~\ref{material} the viscoplastic model is presented. A criteria  for the definition of the time steps is proposed, based on the control of the increment of a sensitive variable. Then the reference model is presented in Section~\ref{sec:ref} with focus on the chosen procedure for the determination of the time discretization. In Section~\ref{sec:gloloc} the bases of the noninvasive global/local method  are presented without addressing the problem of the coupling in time. In Section~\ref{global-local time} two strategies for the global/local coupling in time are proposed. Those strategies are compared on the basis of a 2D example. A verification of the whole procedure on a realistic 3D industrial model involving non-matching interfaces is performed in Section~\ref{sec:3D}.

\section{ About the elastoviscoplastic model and its sensitivity to time integration}
\label{material}

\subsection{Material model}
The material model is the one proposed in \cite{LON13}, adapted from the Marquis-Chaboche's behavior \cite{CHA89}. It is devoted to the description of a variety of phenomena which are characteristic of the elastoviscoplastic response under cycling loading. 
The elasticity itself is linear and isotropic, characterized by Young's modulus~$E$ and Poisson's ratio~$\nu$. 

The nonlinear part of the model is ruled by the yield function based on von Mises criterion:
\begin{equation}
	f = J_2\left(\mat{\sigma}_D - \mat{X}_f\right) - R
\end{equation}
	where $J_2$ is the equivalent von Mises stress with $\mat{\sigma}_D$ the deviatoric stress. The isotropic hardening is considered{ to be saturated at a value $R_0$, so that $R=R_0+\sigma_y$ where $R_0$ is an offset added to the elastic yield stress $\sigma_y$}. Nevertheless, it is possible to use an exponential law in order to accurately represent the monotonic traction behavior or the cyclic stabilization. 

$\mat{\epsilon}^p$ is the plastic strain tensor and $p$ is the  accumulated plasticity. These latter are split in a fast part $(f)$ and a slow part $(s)$:
\begin{equation}
\dot{\mat{\epsilon}}^p = \dot{\mat{\epsilon}}^p_f + \dot{\mat{\epsilon}}^p_s
\quad and \quad
p_i = \int_{0}^{t} \sqrt{\frac{3}{2}\dot{\mat{\epsilon}}^p_i : \dot{\mat{\epsilon}}^p_i} {\color{black}d\tau}\quad \text{ with for }i\in\{f,s\}
\end{equation}
The kinematic hardening $X=X_f$ follows the classic Armstrong-Frederick's formulation, only related to the fast potential:
\begin{equation}
\dot{\mat{X}}_f = \frac{2}{3} C \dot{\mat{\epsilon}}^p_f - D \mat{X}_f\dot{p}_f
\end{equation}
where $C$ and $D$ are two material parameters. 
The fast elastoviscoplastic and slow {viscoplastic} potentials follow Norton-Hoff's laws:
\begin{equation}
\dot{p}_f = \left\langle \frac{J_2(\sigma_D-X_f) - {R}}{K_f}\right\rangle_+^{n_f}
\label{eq:fast}
\end{equation}
\begin{equation}
\dot{p}_s = \left\langle \frac{J_2(\sigma_D)}{K_s}\right\rangle_+^{n_s}
\label{eq:slow}
\end{equation}

\begin{remark}
		Even if the slow potential has no elastic threshold, it is a viscoplastic potential in its own right: it leads to plastic strains which remain after a loading/unloading cycle contrary to the case of a viscoelastic potential.
\end{remark}

These constitutive relations allow for phenomena related to high temperature such as creep, strain rate effect, mean stress relaxation, and for all asymptotic behaviors to be represented.
 
The fast potential \eqref{eq:fast} dominates for strain rates in the range  $\left[ 10^{-5},  \;10^{-2}\right]s^{-1}$ whereas the slow one \eqref{eq:slow} governs in the range $\left[ 10^{-9},  \;10^{-5}\right]s^{-1}$. The material of interest being confidential,  we illustrate the type of constitutive response  using the parameters given in \cite{LEM09} for a nickel based superalloy IN100 evolving at $800^\circ$C which is the mean temperature value during a flight. The corresponding values are reported in table \ref{paramMaterial}. 
 
\begin{table}[ht]\centering
	\begin{tabular}{|c|c|c|c|c|c|c|c|c|}
		\hline 
		E [MPa] & $\nu$  & C [MPa]  & $D$  & R [MPa] & $n_f$&$K_f$ & $n_s$&$K_s$  \\ 
		\hline 
		154,000  & 0.28 & 615,000 & 1,870 & 80 & 14 & 630 & 17.2 & 1,300  \\ 
		\hline 
	\end{tabular} 
	\captionof{table}{Material parameters ($800^{\circ}$) from \cite{LEM09}}
	\label{paramMaterial}
\end{table}

In this study, the variation of temperature is taken into account only by making use of the adapted values of the material parameters. Nevertheless, some additional works were carried out with more complex laws involving thermal strains, without any trouble. For confidentiality reasons, we use classical values for the fast potential parameters, while the values for the parameters of the slow potential are chosen in order to balance the two  plastic potentials for a strain rate of $10^{-5}s^{-1}$. The slow potential having no threshold, nonlinearity occurs as soon as the structure is loaded. 

This law is incorporated in \textit{Abaqus Standard} by the \textit{Zmat} tool from \textit{Zset} software \cite{ZSE15}.

\subsection{Sensitivity to the time step: choice of a control parameter} 
\label{sec:autotime}

The reference solver for this kind of nonlinear problem is the incremental Newton-Raphson method. Given a curve of load amplitude as Figure~\ref{fig:cycle}, it is thus important to determine the load increment which ensures a good compromise between precision and CPU time. Engineering rules often permit to determine an efficient prediscretization. Anyhow it is convenient to adapt the next increment to the current properties of the computation, in particular if the solver has trouble to converge; this procedure is called cutback. 

By default, Abaqus adapts the increments based on the monitoring of the speed of convergence, estimated by the evolution of the norm of the residual, compared to theoretical results and heuristics. The current increment being given, if divergence occurs then the computation is restarted with a 4 time smaller increment; if after a certain number of iterations the residual decreases too slowly, the computation is restarted with a 2 time smaller increment. If convergence is fast for two consecutive time steps, the next increment is enlarged by 50\%. 
{\color{black} Abaqus also monitors the precision of the computation. In implicit dynamics analysis, the relevance of the integration is estimated by evaluating the residual in an interpolated mid-increment configuration. In quasi-static analysis, which is our case of interest, it proposes to adapt the time step in order not to allow too large evolution of internal variables.} Note that the cutback procedure is highly customizable, but this process is highly dependent on the model especially for complex structures and it is difficult to set it up {a priori} with efficiency. 

{\color{black}For our problem of interest, it appears that with default settings, cutbacks are only triggered out of convergence issues, and in fact accuracy is not sufficiently monitored. } This is particularly problematic since it is crucial to have comparable levels of precision in the estimation of the inelastic evolution of several models.
\color{black}
\medskip

{\color{black} It is thus critical to carefully tune the precision control of Abaqus which is also available through the \textit{Zmat} tool of the \textit{Zset} software \cite{ZSE15}:} any variable of the constitutive law can be used to master the time refinement based on a given increment of the chosen variable.  After some trials, the fast accumulated plasticity was chosen as control variable {since it appeared that small increments were correlated with high precision}. In what follows, ${\Delta p_{\max}}$ denotes the associated plastic increment threshold: {if a given increment leads to an increase of the accumulated plastic strain $\Delta p$ larger than ${\Delta p_{\max}}$ at any Gauss point then a cutback is applied: the computation is restarted with an increment of size reduced in proportion with ${\Delta p_{\max}}/\Delta p$. Please note that this process is totally automatic for the user, making it very simple to use and to calibrate on an elementary test as explained below. Such time steps added because of accuracy consideration will be called ``additional time steps'' (we reserve the name cutbacks to time steps added based on convergence consideration).}

\begin{figure}[ht]
	\subfloat[Strain rate of $10^{-5}$  \label{fig:inteVite-5}]{
	\includegraphics[width=0.49\textwidth]{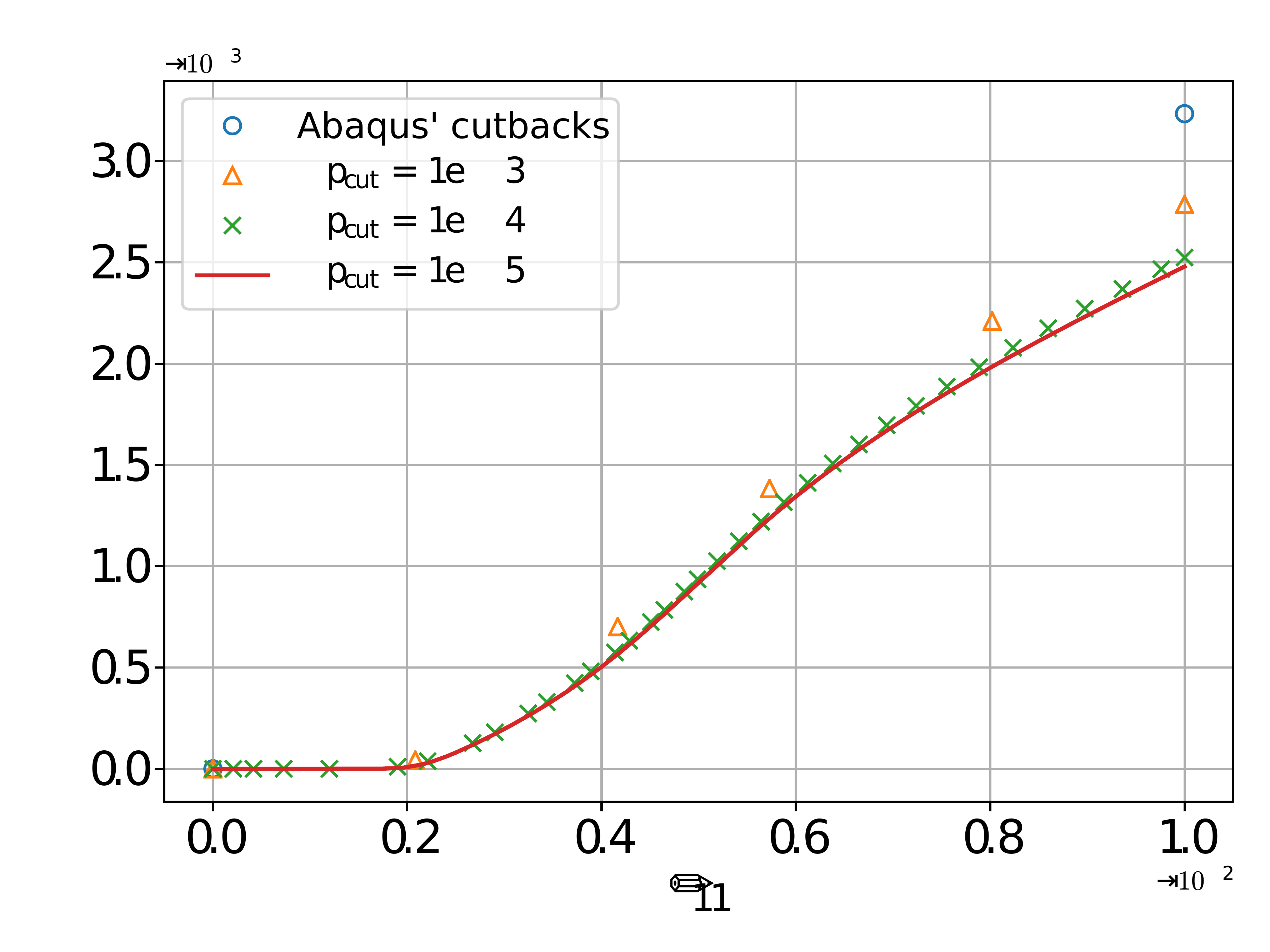}
	}
	\hfill
	\subfloat[Strain rate of $10^{-8}$ \label{fig:inteVit-8}]{
		\includegraphics[width=0.49\textwidth]{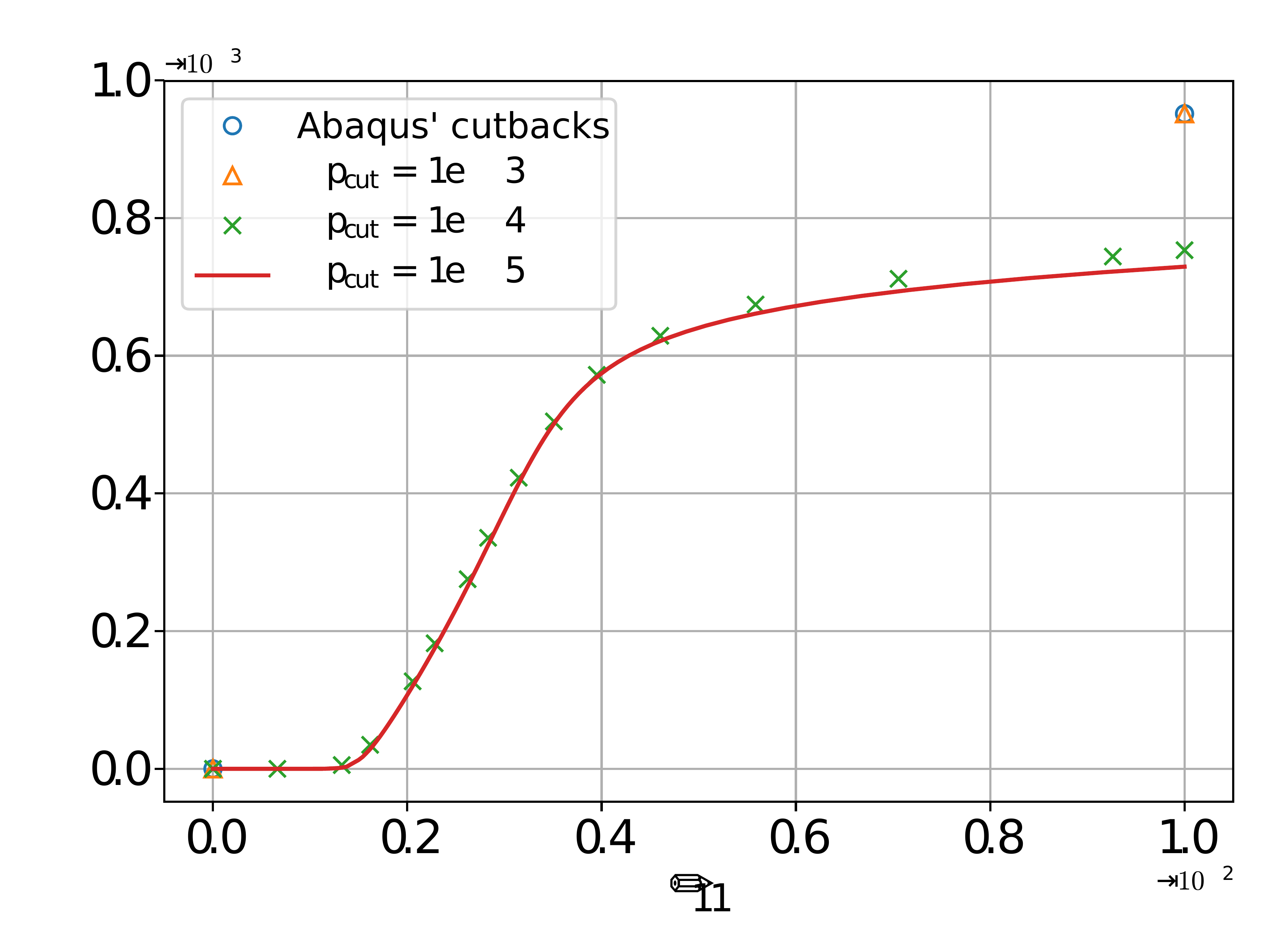}
	}
	\caption{Integration: influence of the time step for IN100 }
	\label{fig:integration}
\end{figure}

Sensible ranges for  ${\Delta p_{\max}}$ can be determined on a tension test  on  a single cubic element for different strain rates, see Figure~\ref{fig:integration}. In Table~\ref{tab:influStrainRate}, the results for different thresholds are compared to an overkill solution obtained with the increment ${\Delta p_{\max}}=10 ^{-5}$. The results obtained using the default cutback procedure of Abaqus, starting from a single initial time increment are also displayed. { Note that in practice for $\Delta p_{\max}\leqslant 10^{-4}$ no cutbacks were triggered out of convergence issues.}

\begin{table}[ht]\centering
	\begin{tabular}{|c||c|c|c|c|}
		\hline 
		Strain rates $\left[s^{-1}\right]$ & { Abaqus default cutbacks } & ${\Delta p_{\max}}=10^{-3}$  & ${\Delta p_{\max}}=10^{-4}$  \\ 
		\hline 
		$10^{-3}$  & 8.81$\%$ (2) & 1.84$\%$ (9) & 0.20 $\%$ (65) \\ 
		\hline 
		$10^{-5}$  & 30.40$\%$ (2) &12.36$\%$ (6) & 1.80$\%$ (36) \\
		\hline 
		$10^{-8}$  & 30.46$\%$ (2) & 30.46 $\%$ (2)& 3.28$\%$  (17)  \\
		\hline 
	\end{tabular} 
	\captionof{table}{Error levels for different strain rates: influence of the threshold and results obtained with Abaqus default cutback procedure (with number of time steps) to reach 1\% deformation.}
	\label{tab:influStrainRate}
\end{table}

We observe that the solution obtained with ${\Delta p_{\max}}=10 ^{-4}$ is very close to the reference obtained with ${\Delta p_{\max}}=10 ^{-5}$, at a much lower computational cost (65 time steps instead of 500, in the case of fast loading). Note that, all errors are measured for a strain of $1\%$, which is unlikely to be reached in practice at low strain rates, so that the lower precision observed with low strain rates (3.3\%  for $\dot{\varepsilon}=10^{-8}s^{-1}$ vs 0.2\% for $\dot{\varepsilon}=10^{-3}s^{-1}$) does not lead to loss of accuracy on structural applications.

\section{Application to the reference 2D model} 
\label{sec:ref}

%Before considering local/global coupling, we analyze how the increment adaptation procedure behaves on 2D monolithic problem which approach remains computable in a reasonable duration.	
{Before considering local/global coupling, we analyze how the increment adaptation procedure behaves on a reference 2D problem which is sufficiently small to be solved in a reasonable time using both monolithic and global/local approaches.}

\subsection{Reference monolithic model}

\begin{figure}[ht]
	\subfloat[Mesh, external loads and boundary conditions \label{fig:Ref-Mesh}]{
		\includegraphics[width=0.49\textwidth]{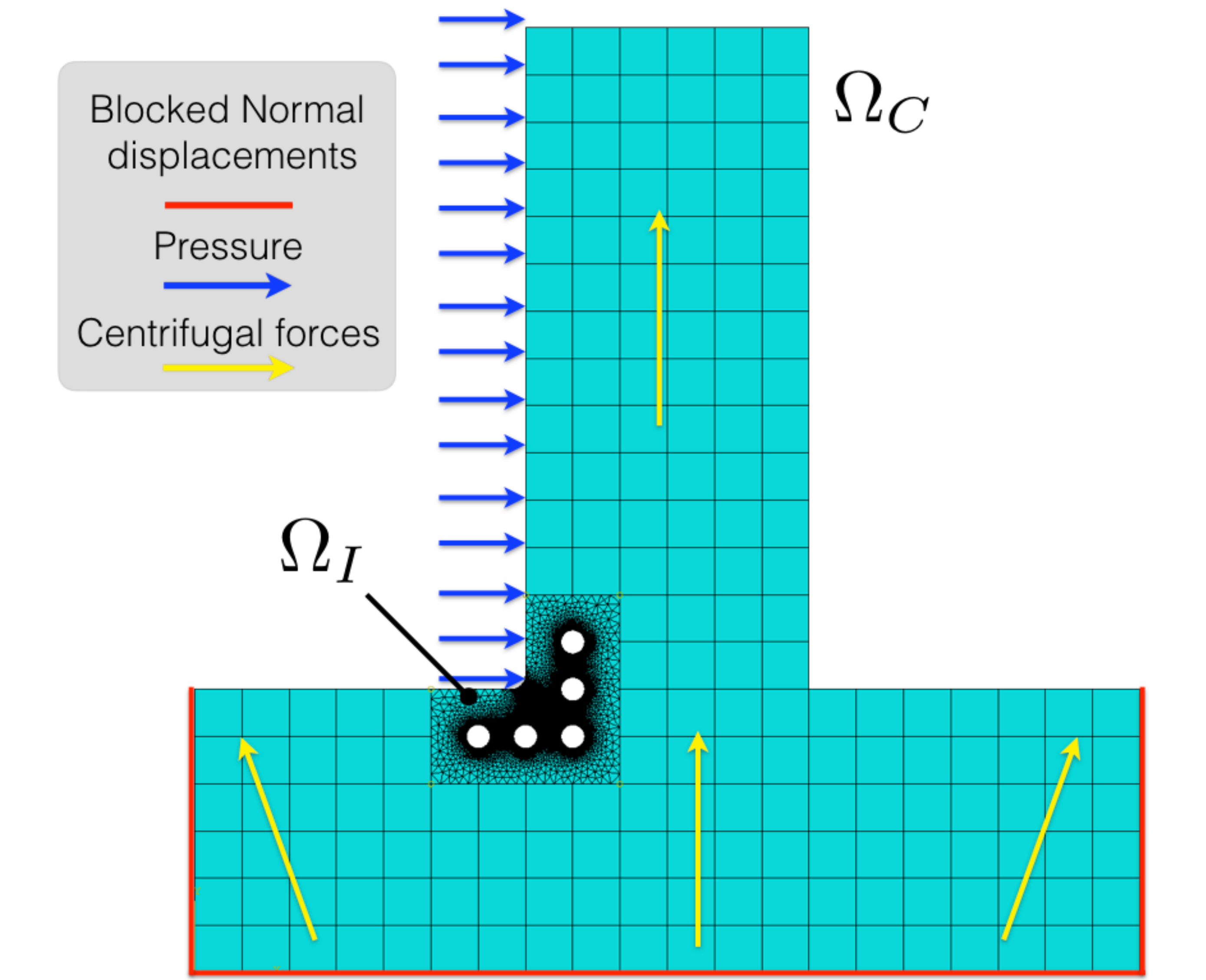}
	}
	\hfill
	\subfloat[Cycle definition \label{fig:cycle}]{
		\includegraphics[width=0.49\textwidth]{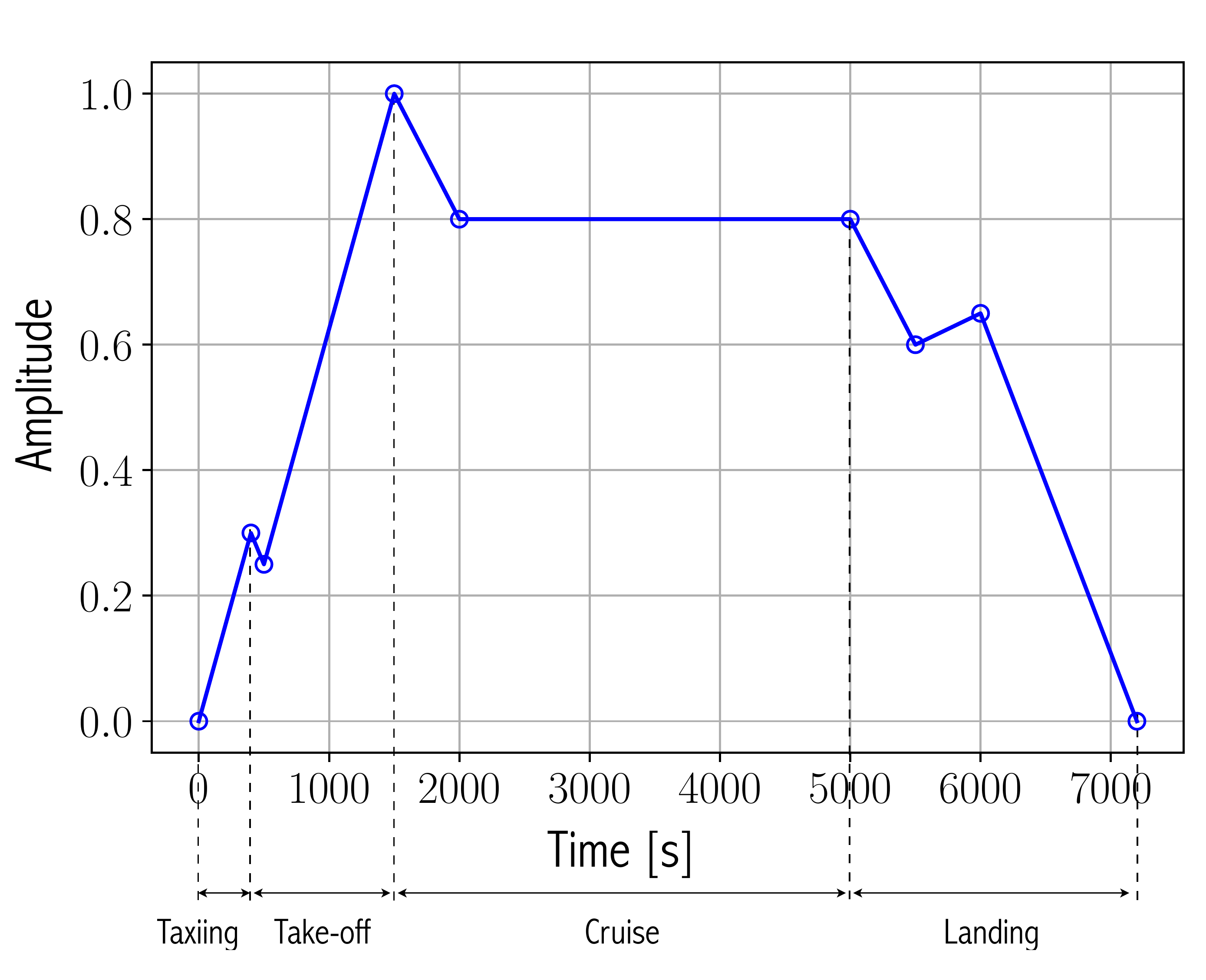}
	}
	\caption{Reference 2D model}
	\label{fig:referenceModel}
\end{figure}

The reference monolithic model is the target of the analysis; it is computed in order to evaluate the accuracy of the global/local method. It represents the structure with all the geometrical details, see Figure~\ref{fig:referenceModel}. The normal displacement of the foot of the part is prescribed to be zero, the air flux on the leading edge is modeled by a constant pressure and the rotation of the blade is taken into account by a centrifugal body force. The intensity of these external loads evolves in time according to the cycle of Figure~\ref{fig:referenceModel} which mimics the main phases experienced by an engine during one flight. {\color{black} Note that all geometric nonlinearities are being neglected; in particular, the loads are evaluated in the initial configuration.}

\subsection{Time prediscretization over the cycle}
\label{globaltimediscr}
Clearly, the definition of the load cycle by 8 time increments of Figure~\ref{fig:referenceModel} is far from what is needed to ensure convergence and precision, or in other terms, it is far from the final time discretization resulting from cutbacks and additional time steps. In order to build a fair comparison with the global/local coupling, we propose to incorporate in the initial time discretization all the additional time steps that are deemed necessary for the resolution of the simplified model that will serve as the \emph{global} model of the coupling with given  $\Delta p_{\max}$ (see Subsection~\ref{ss:global}). In our case, the simplified model has the same behavior as the reference, but it bears no geometrical details, it is presented on Figure~\ref{2Dloads}.

%In order to make the time-refinement process simpler, a prediscretization is performed before the computation. It consists in enriching the initial eight-step discretization of the cycle of Figure~\ref{fig:referenceModel}  by the { additional time steps} required by a simplified model without geometrical details for a given ${\Delta p_{\max}}$. This simplified model will serve as the \textit{global} model of the global/local coupling, it is represented on Figure~\ref{2Dloads}.
The time discretization resulting from this prior analysis is called \emph{prediscretization}. It is then used for the reference monolithic solution (with the same $\Delta p_{\max}$). \medskip

Table~\ref{tab:refDiscr} presents the number of time steps of the prediscretization and in the reference computation, for the Abaqus default cutback strategy and for different values of $\Delta p_{\max}$. It also gives the time spent for the reference computation. We observe that Abaqus default cutbacks introduce no new time step in the prediscretization and only 4 more for the reference computation. The control on plasticity increment leads to introducing more time steps in the prediscretization (up to 36 added time steps for $\Delta p_{\max}=10^{-4}$ or $10^{-5}$), the value of $\Delta p_{\max}$ is much more influent during the reference computation where, roughly, one order of magnitude gained in precision corresponds to the increase by one order of magnitude of the number of time steps (and of the CPU time).

\begin{table}[ht]
	\begin{center}
		\begin{tabular}{|c||c|c|c|}
			\hline 
			Integration technique & Prediscr. \#steps & Final \#steps & Total time \\ 
			\hline
			\hline 
			{Abaqus default cutbacks} & 8 & 12 & 1m 11s \\
			\hline 
			${\Delta p_{\max}}=10^{-3}$ & 12 & 27   & 1m 33s   \\ 
			\hline 
			${\Delta p_{\max}}=10^{-4}$ & 44 & 173  & 7m 11s   \\ 
			\hline 
			${\Delta p_{\max}}=10^{-5}$ & 44 & 1501 & 1h 1m 29s \\
			\hline 
		\end{tabular} 
	\end{center}
	\caption{Influence of time integration on the computation (2D monolithic model)}
	\label{tab:refDiscr}
\end{table}

\begin{figure}[ht]
	\subfloat[Prediscretization (global computation) \label{fig:globPreDiscr}]{
		\includegraphics[width=0.49\textwidth]{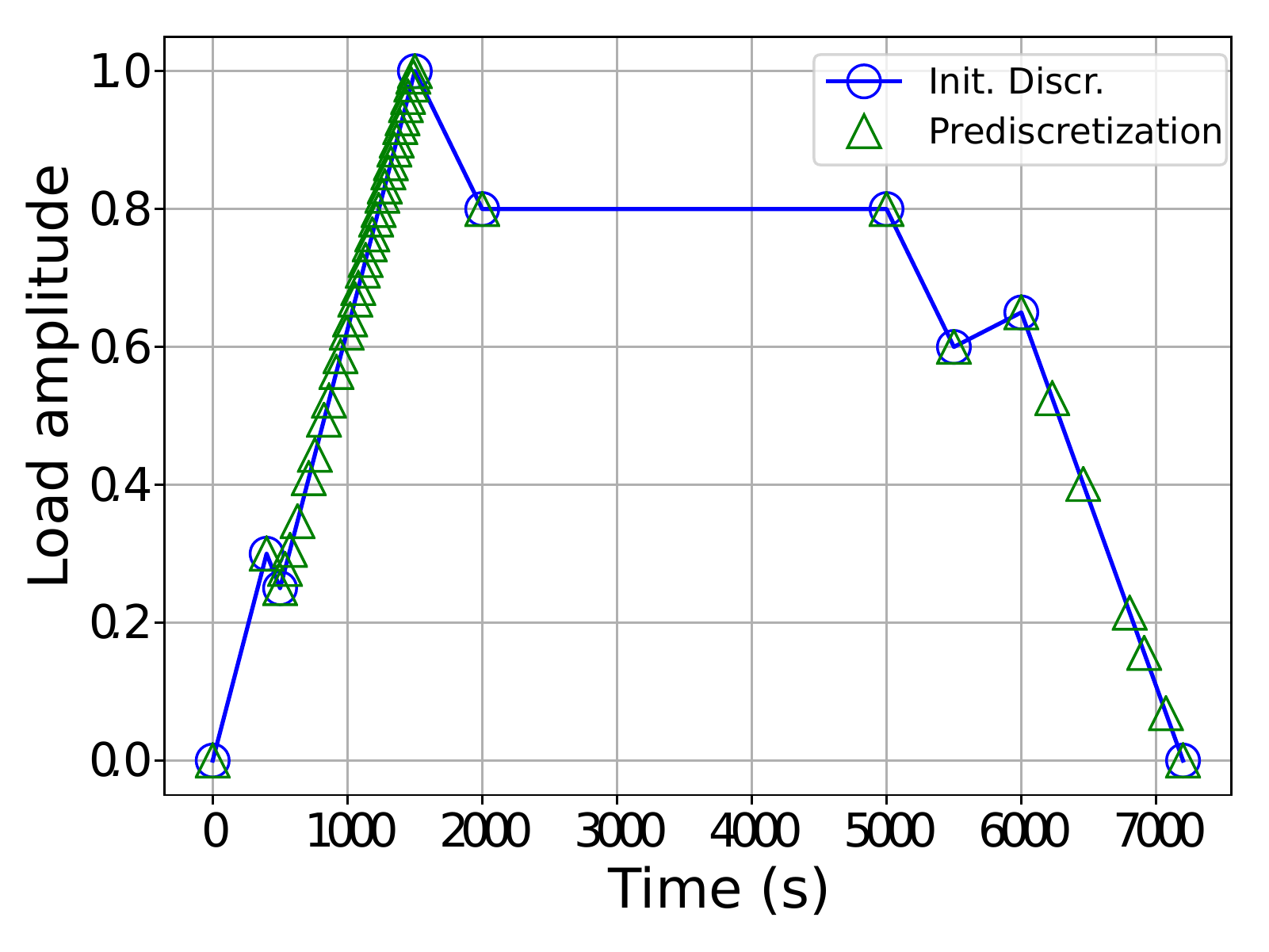}
	}
	\hfill
	\subfloat[Reference time grid \label{fig:refInte1e-4}]{
		\includegraphics[width=0.49\textwidth]{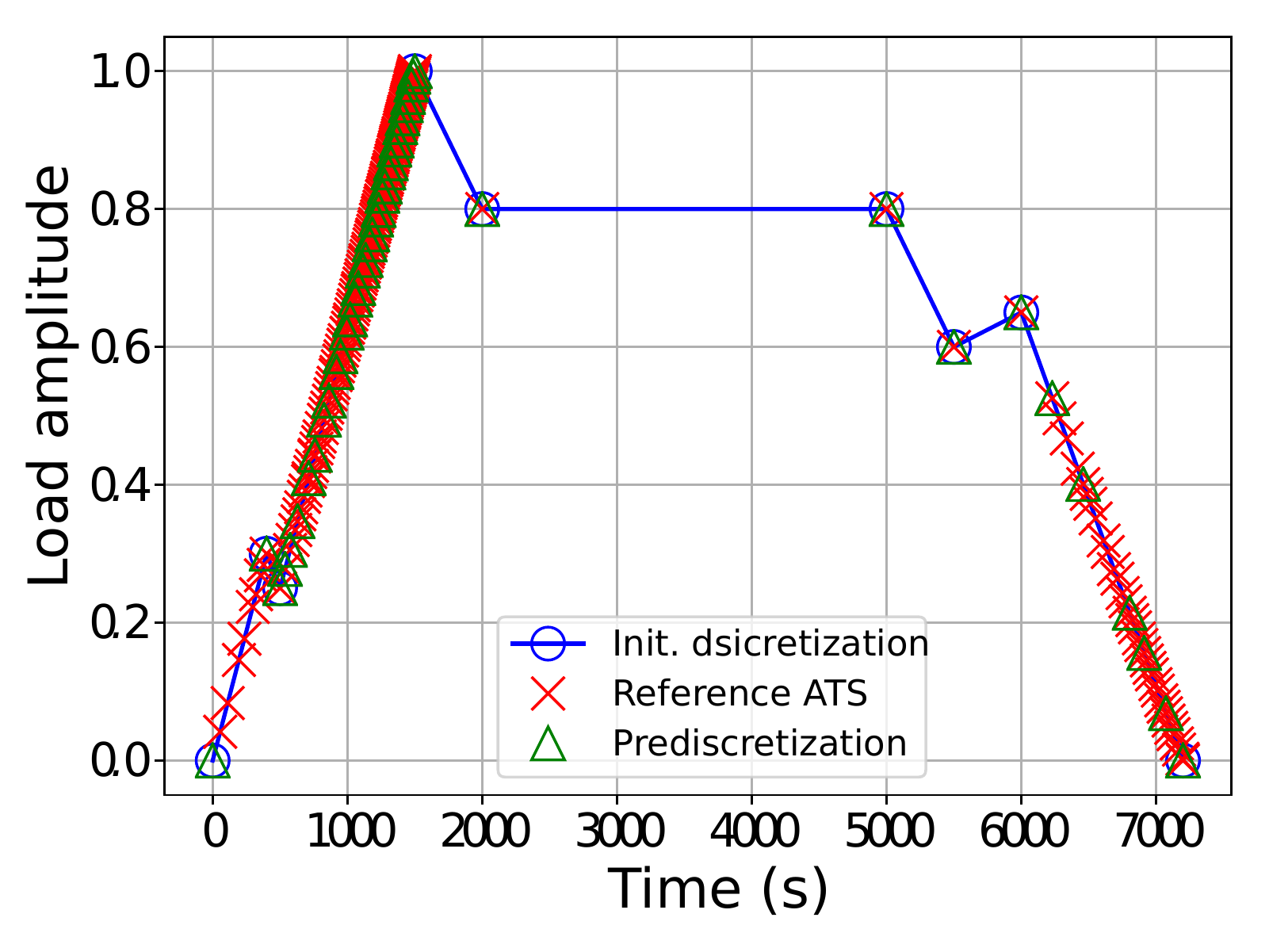}
	}
	\caption{Time grids for the global and reference models (${\Delta p_{\max}} = 10^{-4}$)}
	\label{fig:timeIntegrationGR}
\end{figure}

Figure~\ref{fig:timeIntegrationGR} presents the grids resulting from the prediscretization and from the reference computation. Not surprisingly, we observe that added time steps are mostly located in the phase of strongly increasing load, and to a lesser extent in the strongly unloading phase. 

\color{black}

\subsection{Accuracy of the computations}

%{Results presented in Table~\ref{tab:maxElemPrecRef} show that the calibration on an elementary model leads to an equivalent accuracy at the structural level. It can be considered as a preliminary stage of the analysis guarantying the constitutive law to be well integrated.}

\begin{table}[ht]
	\begin{center}
		\begin{tabular}{|c||c|c|c||c|}
			\hline 
			& ${\Delta p_{\max}}\!=\!10^{-5}$	&{Abaqus default cutbacks} & {${\Delta p_{\max}}\!=\!10^{-3}$} & {${\Delta p_{\max}}\!=\!10^{-4}$}   \\ 
			\hline 
			& Ref. Values	&  Error $\left[\%\right]$ & Error $\left[\%\right]$& Error $\left[\%\right]$ \\ 
			\hline \hline 
			Mises & 650.9 MPa & 10.9 & 1.64 & 0.159  \\ 
			\hline 
			$p_f$&	$1.05\, 10^{-2}$ &  18.4  & 5.05 & 0.871 \\ 
			\hline 
			$p_s$& $1.27\, 10^{-2}$ 	& 3.97 &  1.90  & 0.277 \\ 
			\hline
		\end{tabular} 
	\end{center}
	\caption{Error levels on the most loaded element at the end of the cycle (2D monolithic model)}
	\label{tab:maxElemPrecRef}
\end{table}

The results obtained for several thresholds are compared both in term of accuracy and cost. Due to the results of Section~\ref{material}, ${\Delta p_{\max}}=10^{-5}$ serves to compute the reference solution and to measure errors. Two indicators are used, the von Mises stress and the cumulated plasticity.
%\begin{equation}
%\left\lbrace
%\begin{aligned}
%\eta_{X} &=  \frac{\| X^{GL}  -  X^{R} \|_2 }{\| X^R\|_2 } \\
%\Delta X &= \| X^{GL}  -  X^{R} \|_2
%\end{aligned}
%\right.
%\end{equation}

As observed in Section~\ref{material}, we see in Table~\ref{tab:maxElemPrecRef} that the results obtained with ${\Delta p_{\max}}=10^{-4}$ are very close to the ones obtained with ${\Delta p_{\max}}=10^{-5}$. Table~\ref{tab:refDiscr} shows that using ${\Delta p_{\max}}=10^{-4}$ leads to a much shorter computational time. A duration of one hour for such a simple case would correspond to very long computation time on relevant 3D industrial examples, as the one treated at the end of the paper. That is why we consider  ${\Delta p_{\max}}=10^{-4}$ as a sensible compromise between accuracy and computational time.  

\begin{figure}[ht]
	\centering
	\includegraphics[width=0.95 \textwidth]{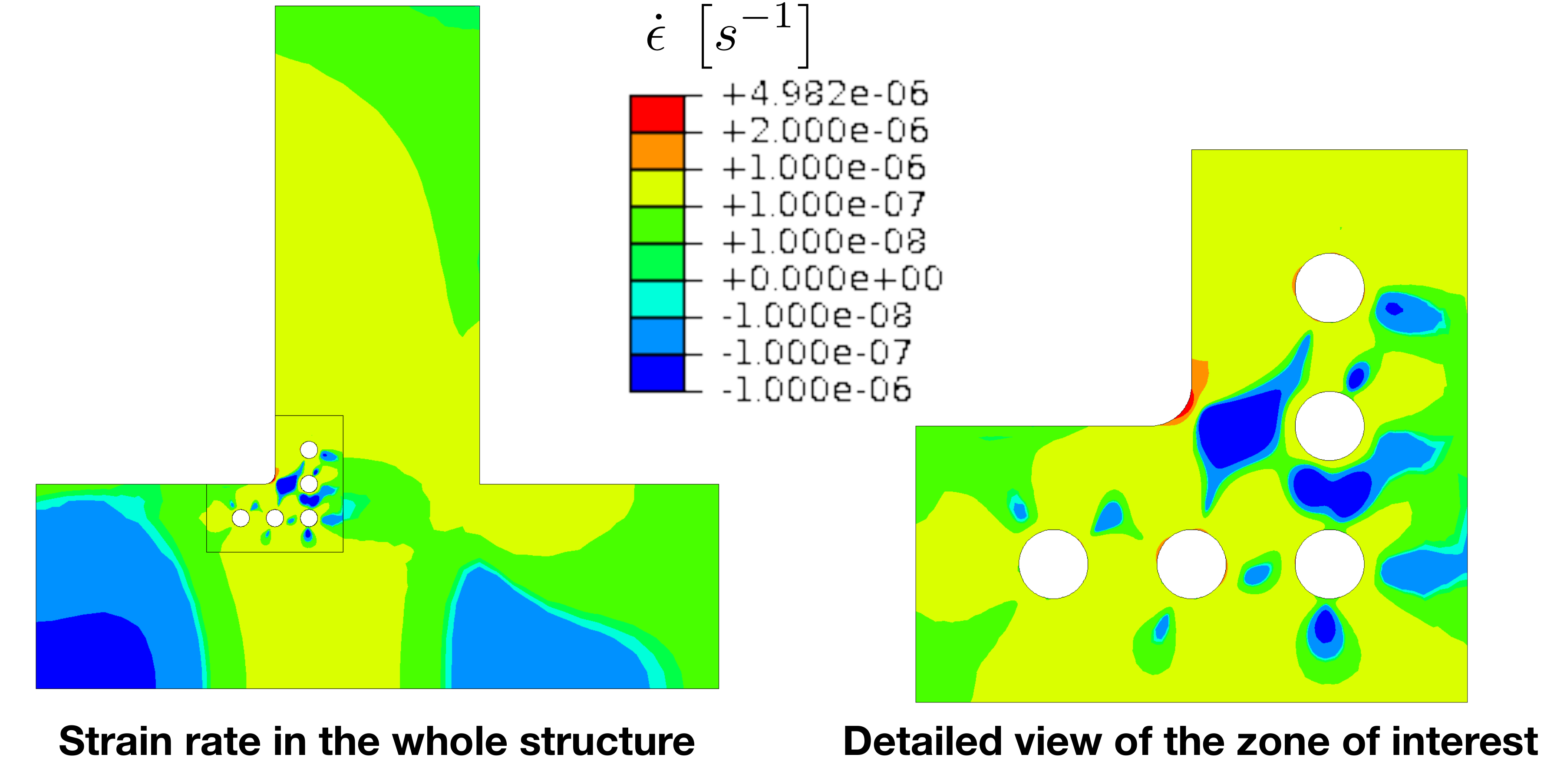}
	\caption{Reference monolithic computation: generation of heterogeneous strain rates in the structure (unloading phase)}
	\label{fig:structStrainRates}
\end{figure}

Let us note that the error indicators are lower than the one obtained in the homogeneous case of Section~\ref{material}, see Table~\ref{tab:influStrainRate}. As said earlier, the strain rate heterogeneity (see Figure~\ref{fig:structStrainRates}) leads to strain heterogeneity, so that error prone situations of large strain reached at low strain rate are never met in practice.

%\begin{figure}[ht]
%	\subfloat[Strain rate at the Global scale \label{fig:SR-Global}]{
%		\includegraphics[width=0.49\textwidth]{figure_4a}
%	}
%	\hfill
%	\subfloat[Zoom in the zone of interest \label{fig:StimeIntegrationGRR-Local}]{
%		\includegraphics[width=0.49\textwidth]{figure_4b}
%	}
%	\caption{Reference monolithic computation: generation of heterogeneous strain rates in the structure (unloading phase)}
%	\label{fig:structStrainRates}
%\end{figure}

\section{Non-invasive global/local method for local and global nonlinear models}
\label{sec:gloloc}
In this section, the basic aspects of the global/local coupling are presented as if the local and global time grids were identical. The treatment of different time discretizations is discussed in Section~\ref{global-local time}.

\begin{figure}[ht]
\centering
\includegraphics[width=1.\textwidth]{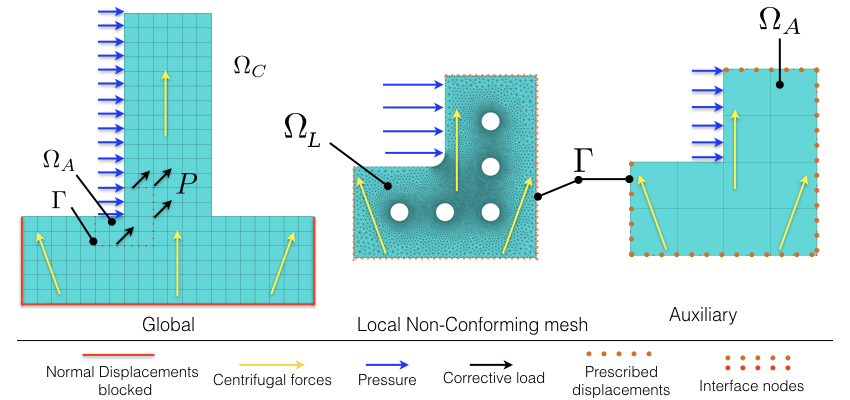}\caption{Models and external loads set up}
\label{2Dloads}
\end{figure}

The strategy makes use of three computational models based on the decomposition of the reference problem as explained in \cite{GOS17} in details for such nonlinear framework. The global model (index $G$) represents the whole structure and may bear a rough description of the zone of interest. The restriction of the global model on the zone of interest is what we call an auxiliary model (index $A$), the complement to the auxiliary model in the global model is written with the index~$C$. The local model (index $L$) bears all the complexity of the zone of interest.
Figure~\ref{2Dloads} illustrates the models used to set up the algorithm.

The global/local ($GL$) solution is defined on the reference model ($\Omega_R=\Omega_L\cup\Omega_C$) as the replacement of the global solution by the local one in the zone of interest:
\begin{equation}
\left( \vec{u}^{GL},\;\mat{\sigma}^{GL}\right) = \left\lbrace 
\begin{aligned}
\left( \vec{u}^{L},\;\mat{\sigma}^{L}\right)  & \; \text{in} \; \Omega_L \\
\left( \vec{u}^{G},\;\mat{\sigma}^{G}\right)   & \; \text{in} \; \Omega_C \\
\end{aligned}
\right.
\end{equation}

\subsection{Global model}\label{ss:global}
The global model matches the reference model on the complementary zone, that is to say everywhere except in the zone of interest where a coarser representation (named auxiliary model) can be used. 
The principle of the algorithm  is to find  an extra load $P$ to be applied on the interface $\Gamma$ of the global model between the complement and the auxiliary zones, such that the local and complement models are in balance at the interface. This is done iteratively by correcting  $\vec{P}$  knowing the difference between the nodal reactions of the complement and local models. For a current value of $P$  the finite element global problem can be written as:
\begin{equation}\text{Given }P,\text{ find }\vec{u}^G=(\vec{u}^G_C,\vec{u}^G_{\Gamma},\vec{u}^G_A)\text{ such that }
\begin{bmatrix}
\vec{f}^G_{int,C}(\vec{u}^G_C,\vec{u}^G_{\Gamma}) \\
\vec{f}^G_{int,\Gamma}(\vec{u}^G_C,\vec{u}^G_{\Gamma},\vec{u}^G_A)\\
\vec{f}^G_{int,A}(\vec{u}^G_{\Gamma},\vec{u}^G_A)
\end{bmatrix}
 + 
\begin{bmatrix}
\vec{f}^G_{ext,C}\\
\vec{f}^G_{ext,\Gamma}+\vec{P}\\
\vec{f}^G_{ext,A}
\end{bmatrix}
= \vec{0}
\end{equation}

For the determination of the different nodal reactions on the interface some codes allow the shape function on one subdomain to be integrated, leading to the following formula for the nodal reaction $\lambda^X_n$ at one node $n\in\Gamma$ seen from domain $\Omega_X$ ($X\in\{C,A,L\}$):
\begin{equation}\label{eq:reacForces}
\begin{aligned}
\lambda^X_n&=\int\limits_{\Omega_X} \left( \sigma_h:\epsilon(\phi_n)-f\cdot \phi_n \right)d\Omega \; -\; \int\limits_{\partial_F\Omega_X} F\cdot \phi_n\;dS
\end{aligned}
\end{equation}
where  $\sigma_h$ denotes the finite element approximation of $\sigma$, $f$ is the given body force and $F$ the given traction on $\partial_F\Omega_X$; $\phi_n$ is the shape function associated with node $n$.

When such a computation is not possible or too difficult (as it is the case using Abaqus),  we make use of the auxiliary model. $\lambda^A$ is then post-processed from a computation on $\Omega^A$ with imposed Dirichlet conditions corresponding to the value of the current global displacement on~$\Gamma$:
\begin{equation}
\text{Given }u^G_{\Gamma},\text{ find }(\lambda^A,\vec{u}^A)\text{ such that }
\begin{bmatrix}
\vec{f}^A_{int,\Gamma}(\vec{u}^A;\vec{u}^G_{\Gamma})\\
\vec{f}^A_{int,A}(\vec{u}^A;\vec{u}^G_{\Gamma})
\end{bmatrix}
+ 
\begin{bmatrix}
\vec{f}_{ext,\Gamma}^A + \vec{\lambda}^A\\
\vec{f}^A_{ext,A}\\
\end{bmatrix}
= \vec{0}
\label{eq:aux}
\end{equation}

\subsection{Local model}
The local model is the restriction of the reference on the zone of interest. The model is computed with the current global displacement prescribed on  $\Gamma$, $u^L_\Gamma := u^G_\Gamma$. The local system thus can be written as:
\begin{equation}
\text{Given }u^G_{\Gamma},\text{ find }(\lambda^L,\vec{u}^L)\text{ such that }
\begin{bmatrix}
\vec{f}^L_{int,\Gamma}(\vec{u}^L;\vec{u}^G_{\Gamma})\\
\vec{f}^L_{int,L}(\vec{u}^L;\vec{u}^G_{\Gamma})
\end{bmatrix}
+ 
\begin{bmatrix}
\vec{f}_{ext,\Gamma}^L + \vec{\lambda}^L\\
\vec{f}^L_{ext,L}\\
\end{bmatrix}
= \vec{0}
\label{eq:local}
\end{equation}
The operation performed here consists in computing the local solution submitted to global displacements (Dirichlet condition) and then extracting corresponding reaction forces $\vec{\lambda}^L$.

\subsection{Non-matching interfaces}
\label{ssec:non-matching}
Handling non-matching interfaces provides flexibility for non-invasive approaches. It facilitates the meshing of the zone of interest  (\textit{cf} Figure \ref{2Dloads}). This procedure appears important especially for complex 3D meshes. Indeed two configurations can be encountered:
\begin{itemize}
\item The zone of interest is detected after a first global computation and then it is defined by a selection of global elements. 
\item The zone of interest can be defined at the level of the CAD. In that case, the local, the complement and the auxiliary models can be meshed independently so that the reference is defined with a {\texttt{TIE} coupling \cite{ABA16} between the master complement and slave local models. The slave nodes are bound to the follow the kinematics of the interface of the complement domain.}

\end{itemize}
In both cases the interface is well defined in all meshes and the coupling can be done on a surface, so that complex integration techniques \cite{GLO05,RUE14} are avoided.

In case of non conforming meshes the coupling equations take the following form:
\begin{equation}
\left\lbrace
\begin{aligned} 
T\vec{u}^G_\Gamma -  \vec{u}^L_\Gamma&=0\\
\vec{\lambda}^G + T^T \vec{\lambda}^L&=0
\end{aligned}\right.
\end{equation}
where $T$ is the (global to local) transfer matrix. In previous studies in the global/local framework, the transfer matrix $T$ was computed with techniques derived from the mortar method \cite{DUV16,LIU14}. Here a simpler solution is used: $T$ is the interpolation matrix which is processed using a \textit{Python} script, before the computation. 

\subsection{Iterations}

The formulation of the algorithm presented in \cite{GEN09a} with a linear global problem is here extended to the case where both the global and local models are nonlinear. For one increment, we have: 
\begin{enumerate}
\setcounter{enumi}{-1}
\item \textbf{Initialization.} Set $i=0$, $\vec{P}_i=\vec{0}$.
\item \textbf{Global computation.} Computation with the interface load $\vec{P}_{i}$, the interface displacements $\vec{u}^G_{\Gamma,i}$ are deduced.
\item \textbf{Local analysis.} (Descent step) Computation with  given external loads under prescribed interface displacements $\mat{T}\vec{u}^G_{\Gamma,i}$. The reaction forces $\vec{\lambda}^{L}_{i}$ are deduced.
\item \textbf{Auxiliary analysis.} Computation with  given external loads and prescribed displacements $\vec{u}^G_{\Gamma,i}$ applied on $\Gamma$. The reaction forces $\vec{\lambda}^A_{i}$ are deduced.
\item \textbf{Computation of the interface equilibrium residual.} 
\begin{equation}
\vec{r}_{i} = -( \mat{T}^T\vec{\lambda}^L_{i}+ \vec{\lambda}^C_{i})=  -( \mat{T}^T\vec{\lambda}^L_{i}- \vec{\lambda}^A_{i}+\vec{P}_{i})
\end{equation}
\item \textbf{Update of $\vec{P}$}:
\begin{equation}\label{eq:update}
\vec{P}_{i+1} = \vec{P}_{i} + \vec{r}_{i} = \vec{\lambda}^A_{\Gamma,i} - \mat{T}^T\vec{\lambda}^L_{i}
\end{equation}
Then $i\leftarrow (i+1)$ and goto 1.
\end{enumerate}

The local and auxiliary analysis are computed in parallel thus the use of the auxiliary model does not add  computational time to the general algorithm.
The algorithm is a stationary iteration; its convergence can be proved under very general hypothesis {(see \cite{NOU17} for a proof with weak hypothesis and \cite{GOS17} for the registration of the method amongst Schwarz alternating methods for which many convergence results exist)}. In the general case, relaxation may have to be used to ensure convergence by modifying \eqref{eq:update} as follows: $\vec{P}_{i+1} = \vec{P}_{i} + \omega\vec{r}_{i}$, with $\omega$ small enough. Relaxation is needed only if the auxiliary model is more compliant than the local model, which is not the case in the studied examples.

\begin{figure}[ht]
	\subfloat[Global/local residual \label{fig:convResidual}]{
		\includegraphics[width=0.49\textwidth]{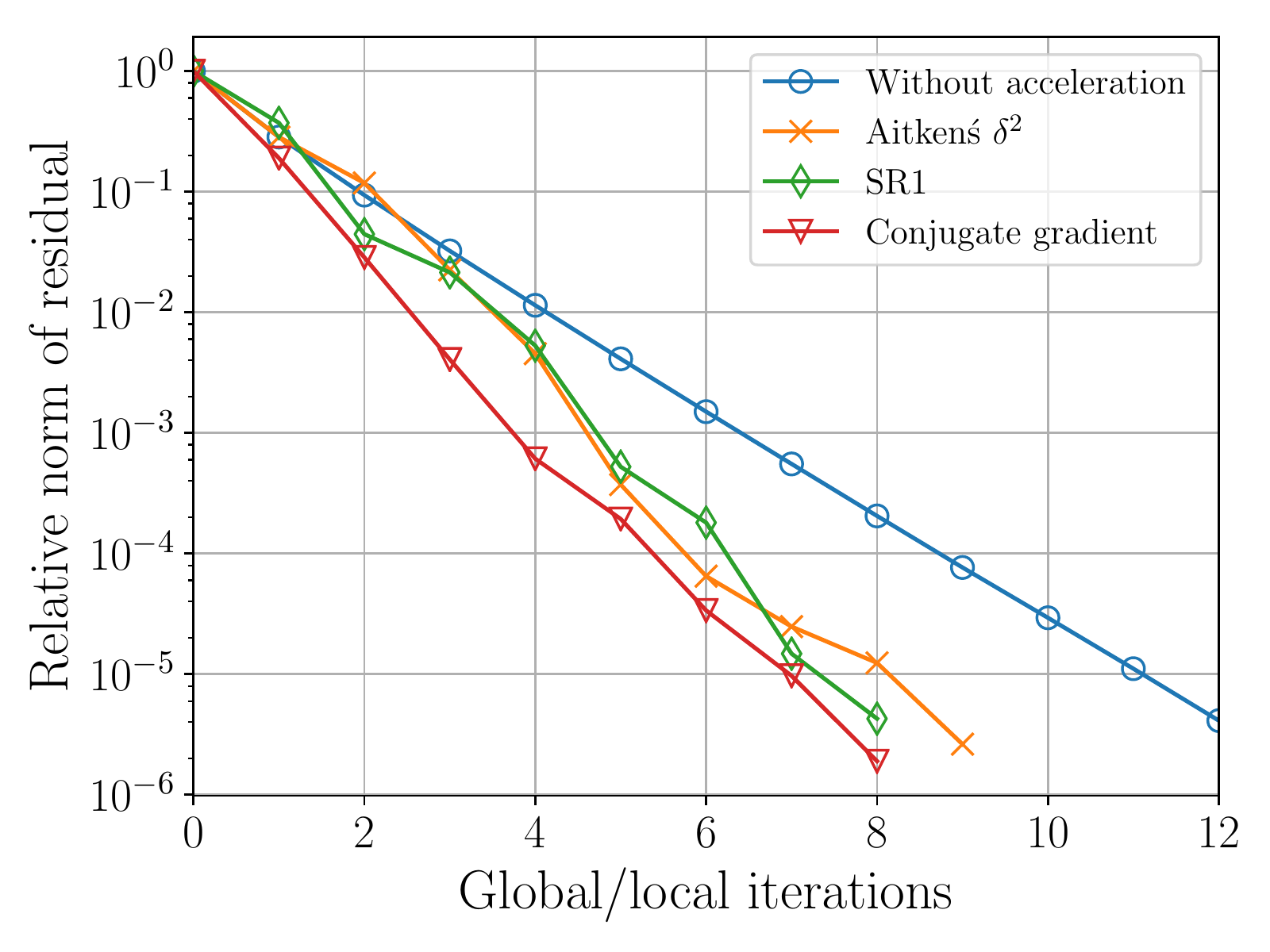}
	}
	\hfill
	\subfloat[True error levels \label{fig:trueErrorP}]{
		\includegraphics[width=0.49\textwidth]{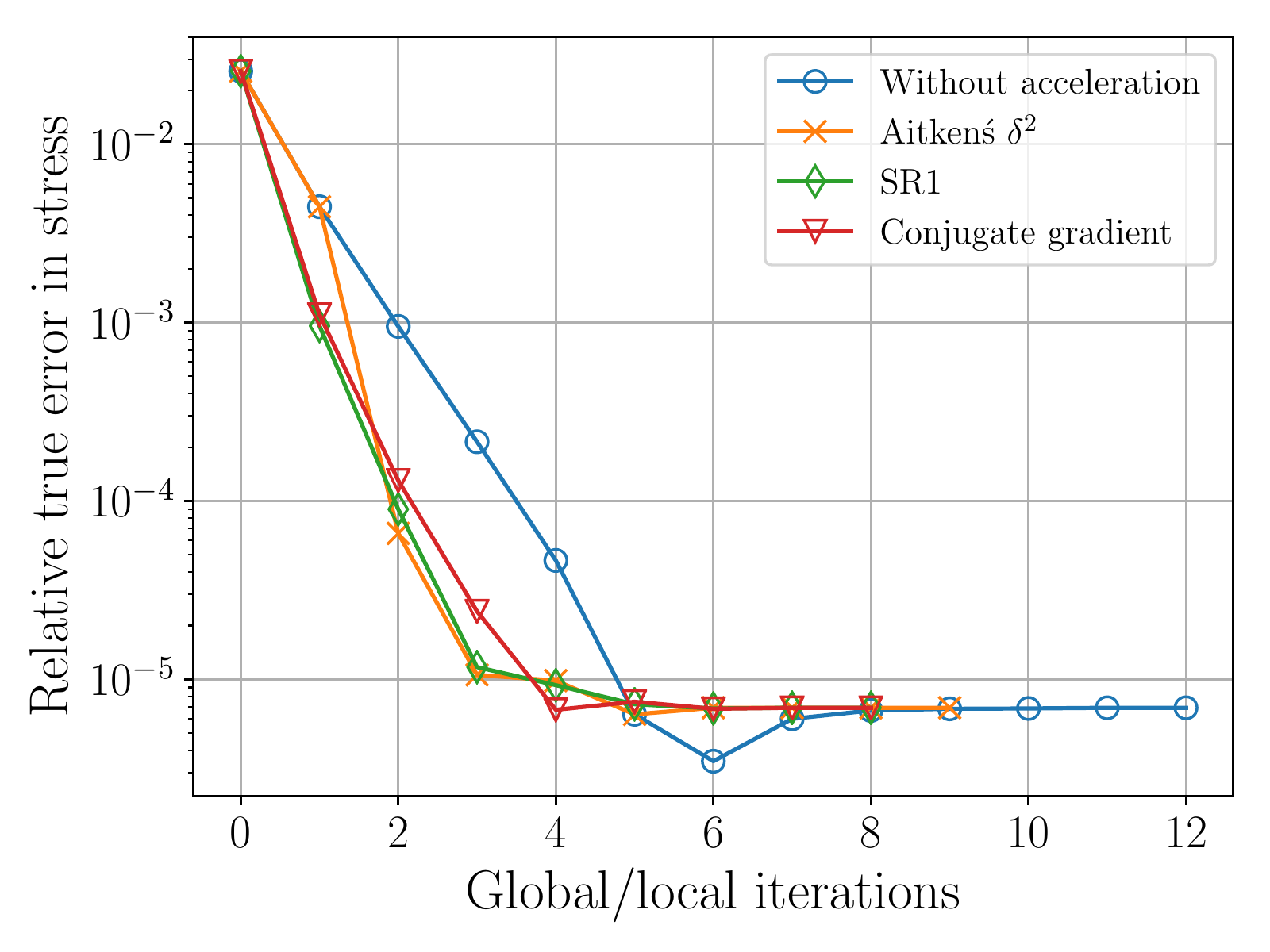}
	}
	\caption{Convergence and acceleration of nonlinear global/local algorithm}
	\label{fig:convergenceAndAccuracy}
\end{figure}

The relative norm of the residual $\|r_i\|_2/\|r_0\|_2$ is used to monitor the convergence of the method over the iterations. {In practice, the tolerance to stop the iterations is set to $10^{-5}$}. Figure~\ref{fig:convergenceAndAccuracy} presents the evolution of the residual and of the true error compared to the monolithic computation, for different acceleration techniques. As usual Aitken's $\delta^2$ leads to significant acceleration at almost null extra cost.

\begin{remark}
The algorithm was presented for one increment of time. In order to compute a loading cycle involving many time steps, one has to reuse the current converged state to initialize the next computation. This is made  using the \texttt{*Restart} function of \textit{Abaqus}.
\end{remark}
\begin{remark}
Stagnation occurs when  observing the true error (with respect to a monolithic computation), as can be seen on Figure~\ref{fig:trueErrorP}. This is problem very commonly encountered when using  \textit{Abaqus/Standard}. {\color{black} A probable cause is that our implementation makes use of output database (odb) files in order to extract the data necessary for the computation of the corrective load to be applied on the global model; and these database truncate nodal reaction to single precision reals in order to limit the size of the files.}
\end{remark}

\section{Time coupling for the Global/Local method}
\label{global-local time}
The main issue, for the considered application, is that the global and local  models  require, in order for them to be properly integrated, different time steps that are not known in advance. The retained principle  here is to make use of the automatic adaptive time stepping control method  presented in Subsection~\ref{sec:autotime}). As explained in Subsection~\ref{globaltimediscr}, we start from the prediscretization which correspond to the grid adapted to the global computation with given threshold $\Delta p_{\max}$. The remaining question is when to apply global/local coupling iterations, in other words at which time steps the models have to be coupled.

{
Let $(t^G_0,t^G_1,\ldots,t^G_N)$ be the initial set of time steps resulting from the prediscretization. They define the initial set of time increments $\Delta t^G_i= t^G_{i+1}- t^G_{i}$.}

\subsection{Couplings strategies}
\label{ssec:autotime}
In what follows two possible coupling strategies are presented. They are compared to the monolithic solution and to the sub-modeling  procedure in terms of accuracy and cost.
\begin{itemize}
\item For the first strategy, called ``weak time coupling'', the coupling is made only at the global additional time steps inserted during the local/global iterations.
\item For the second strategy, called ``full time coupling'', the coupling is ensured at each global and local additional time steps inserted during the iterations.\end{itemize}

\subsubsection{Weak time coupling}\label{ssec:weakCoupling}

\begin{figure}[ht]
	\centering
	\includegraphics[width=1. \textwidth]{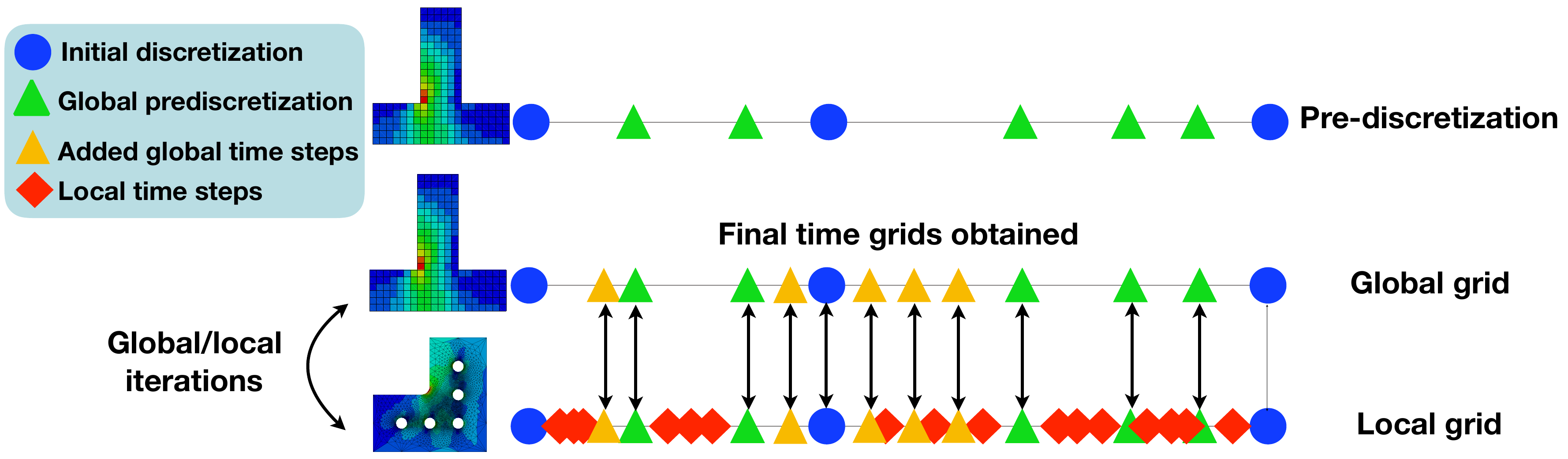}
	\caption{Scheme of weak coupling}
	\label{fig:weakCouplingScheme}
\end{figure}

This approach corresponds to the coupling of the global and local models at all time steps of the global discretization, expecting that the additional time steps, required by the local model, are not of interest for the complementary area. A schematic view of the weak coupling is proposed on Figure~\ref{fig:weakCouplingScheme}. 
\begin{itemize}
	\item Initial cycle definition (blue points)
	\item Global prediscretization (green triangles $(t^G_n)$)
	\item Additional global time steps (orange triangles) which are necessary because the global/local coupling introduces the extra load $P_i$ which increases the level of plasticity.
	\item Additional local time steps (red diamonds) are purely internal to the local computation.
\end{itemize} 

The weak coupling is summed up in Algorithm~\ref{alg:weakCoupling}. The principle is to try to enforce the coupling at a given time step $t_a=t_c+\Delta t_c$ where $t_a$ is the aimed time step, $t_c$ is current converged time step and $\Delta t_c$ is current value of the increment. If at any iteration of the global/local coupling a precision issue is triggered at the global level ($\Delta p^G>\Delta p_{\max}$ for some global computation) then the increment is reduced and $t_a$ redefined; the coupling is restarted. On the contrary, if a difficulty is encountered at the local level, then the local time steps are adapted until $t_a$ is reached; the coupling is not restarted.

\begin{figure}[ht]
	\subfloat[Load \label{fig:addTimeSteps_load}]{
		\includegraphics[width=0.49\textwidth]{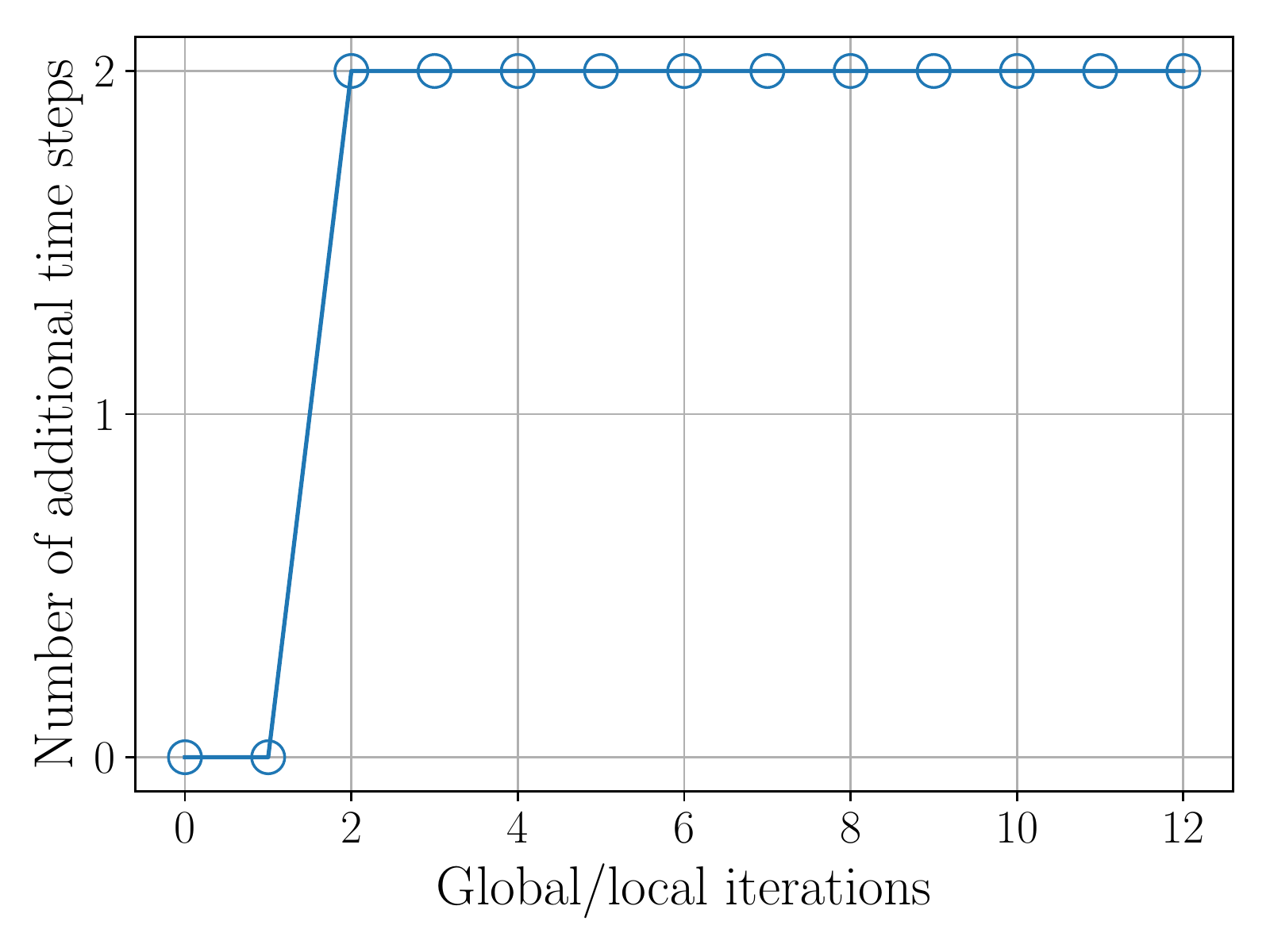}
	}
	\hfill
	\subfloat[Unload \label{fig:addTimeSteps_unload}]{
		\includegraphics[width=0.49\textwidth]{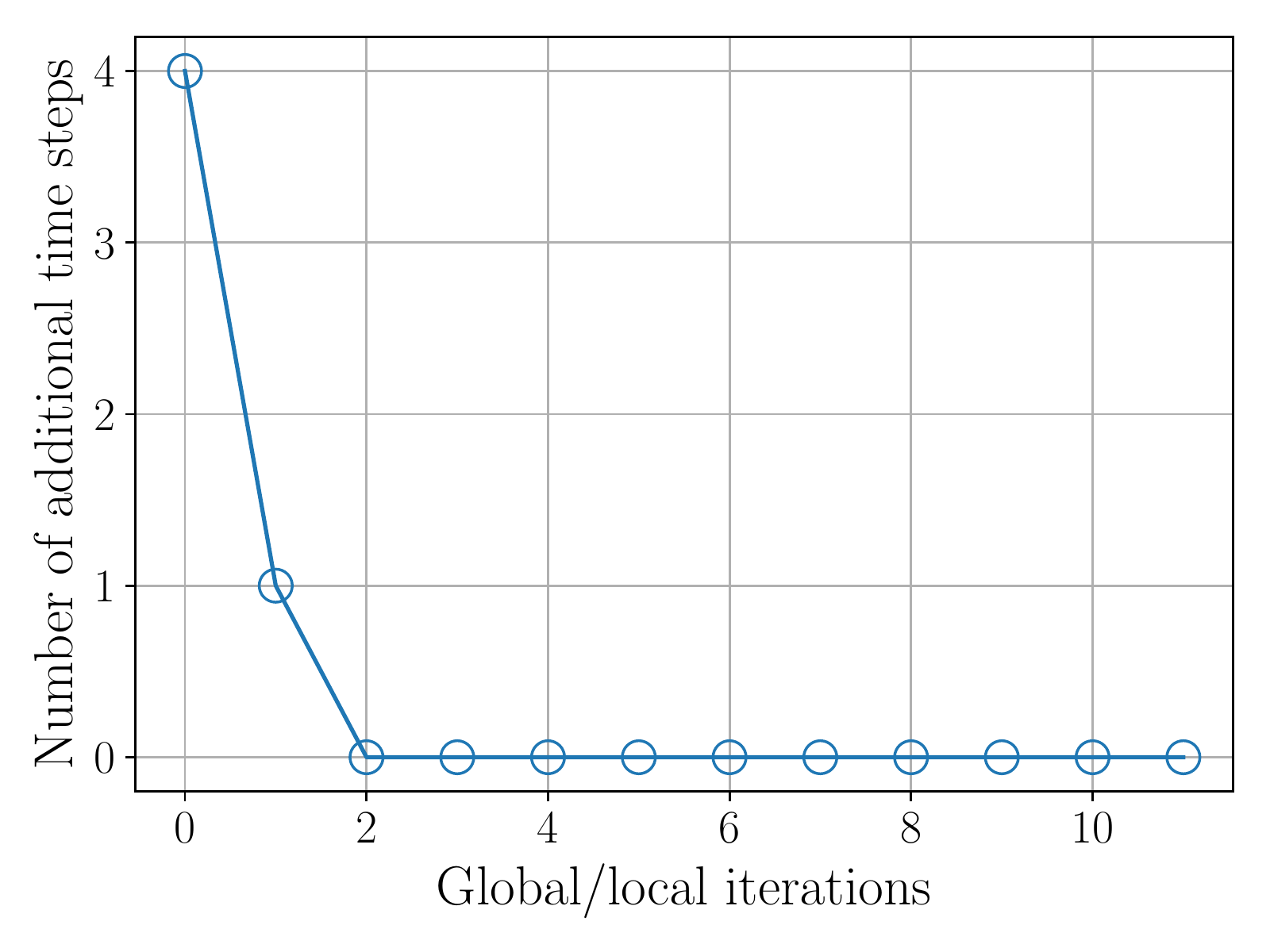}
	}
	\caption{Evolution of \emph{local} additional time steps during global/local iterations in weak coupling}
	\label{fig:addTimeSteps}
\end{figure}

We observed a difference, between loading and unloading phases, regarding the need to insert additional time steps (ATS) at the local scale, depending on the convergence of the coupling, this is illustrated on Figure~\ref{fig:addTimeSteps}. Indeed, during the loading phase, the global model tends to converge ``from below'' and the load transmitted to the local is underestimated in the first coupling iterations, this results in no local ATS in the first two iterations and up to 2 ATS in the following. On the contrary during the unloading phases, plasticity level is overestimated in the first iterations and requires temporarily up to four local ATS whereas once the coupling has almost converged, and $P_i$ is close to the solution, no local ATS are required.

\color{black}

\begin{algorithm2e}[ht]\caption{Weak coupling algorithm }\label{alg:weakCoupling}\DontPrintSemicolon
\tcc{Let $t_n^G$ be the last converge step in the initial grid}
\tcc{Let $t_n^G \leqslant t_c<t^G_{n+1} $ be last converged step}
\tcc{Let $\Delta t_c$ be current time increment}
\KwSty{Aim at} $t_a = \min(t_c+\Delta t_c,t^G_{n+1})$\nllabel{lineWinc}\;
\KwSty{Initialize} Global/local iterations counter: $i = 0$, Load $P_{a,0}$\nllabel{lineWini}\;
\While{$ \|r_i\|_2/\|r_0\|_2 <10^{-5}$}{
		\Begin(Global resolution with extra load $P_{a,i}$){
		\If{$\Delta p^G>\Delta p_{\max}$}{
			\KwSty{Reduce} time increment $\Delta t_c$\; 
			\KwSty{Set} Additional Time Step $t_a \leftarrow t_c + \Delta t_c$\;
			\KwSty{Goto } \ref{lineWini}
		}}
		\Begin(Local and auxiliary resolutions with bc $u^G_{a,i}$){ 
			If needed insert internal local Additional Time steps and interpolate bc \;}
		\KwSty{Evaluate } residual $\|r_{i+1}\|$, \KwSty{Update } $P_{a,i+1}$\;
	    \KwSty{Increment} $i = i + 1 $ \;
		}
	\KwSty{New converged state} $t_c\leftarrow t_a$\;
	\KwSty{Goto } \ref{lineWinc}
\end{algorithm2e}

\subsubsection{Full time coupling} 
\label{ssec:fullCoupling}

The full time coupling consists in performing global/local iterations for all the additional time steps required from both global and local models. It is expected to lead to high level of accuracy at a large computational cost. In Figure~\ref{fig:weakCouplingScheme} the full time coupling would somehow correspond to backporting the local added time steps (red diamonds) to the global grid. Algorithm~\ref{alg:fullCoupling} recaps the full coupling. 

For the full coupling, the unloading phases are particularly critical. Indeed, for the first iterations, the global model provides particularly inadequate boundary condition, resulting in the need of additional time steps for the local model which are no more needed when convergence improves, as evoked for the weak coupling in Figure~\ref{fig:addTimeSteps_unload}. In the full coupling, each additional local time step is transfered on the global time grid and the coupling iteration is restarted. This result in a much increased computational time without practical interest from a precision point of view.

\begin{algorithm2e}[ht]\caption{Full coupling algorithm }\label{alg:fullCoupling}\DontPrintSemicolon
	\tcc{Let $t_n^G$ be the last converge step in the initial grid}
	\tcc{Let $t_n^G \leqslant t_c<t^G_{n+1} $ be last converged step}
	\tcc{Let $\Delta t_c$ be current time increment}
	\KwSty{Aim at} $t_a = \min(t_c+\Delta t_c,t^G_{n+1})$\nllabel{lineFinc}\;
	\KwSty{Initialize} Global/local iterations counter: $i = 0$, Load $P_{a,0}$\nllabel{lineFini}\;
	\While{$ \|r_i\|_2/\|r_0\|_2 <10^{-5}$}{
		{\textbf{Global resolution} with extra load $P_{a,i}$}\;
		\If{$\Delta p^G>\Delta p_{\max}$}{
			\KwSty{Reduce} time increment $\Delta t_c$\; 
			\KwSty{Set} Additional Time Step $t_a \leftarrow t_c + \Delta t_c$\;
			\KwSty{Goto } \ref{lineFini}
		}
		{\textbf{Local and auxiliary resolutions}}\;
		\If{$\Delta p^L>\Delta p_{\max}$}{
			\KwSty{Reduce} time increment $\Delta t_c$\; 
			\KwSty{Set} Additional Time Step $t_a \leftarrow t_c + \Delta t_c$\;
			\KwSty{Goto } \ref{lineFini}
		}
		\KwSty{Evaluate } residual $\|r_{i+1}\|$, \KwSty{Update } $P_{a,i+1}$\;
		\KwSty{Increment} $i = i + 1 $ \;
	}
	\KwSty{New converged state} $t_c\leftarrow t_a$\;
	\KwSty{Goto } \ref{lineFinc}
\end{algorithm2e}
\color{black}

\subsection{Comparison of the different methods in 2D}
\label{ssec:2D}

In this section, submodeling and global/local couplings are compared. 

Table~\ref{tab:inteCycle_couplages} quantifies the additional time steps (ATS) and their impact on CPU time. Global and Local ATS specify which computation did not respect the $\Delta p_{\max}$ criterion and required the insertion of a time step. As explained earlier, the prediscretization was designed so that the global computation of the submodeling approach needs no additional time steps. In the full coupling, the common time discretization is always driven by the local model, which experiences the most severe plastic evolution. In the weak coupling, both global and local models trigger ATS on their own grid, but much less than the full coupling. Regarding the computational time, the full coupling is much too expensive for the objective precision of $\Delta p_{\max}=10^{-4}$.

\begin{table}[ht]
	\begin{center}
		\begin{tabular}{|c||c|c|c|c|c|}
			\hline 
			\multirow{2}{*}{Approaches}	&  \multirow{2}{*}{$\Delta p_{max}$} & Global  & Local  & G/L & Total \\
			&&ATS & ATS & iterations & time \\
			\hline 
			\hline
			\multirow{2}{*}{Submodeling}    & $10^{-3}$ & 0 & 11 & -- &4min \\ 
			\cline{2-6}
			& $10^{-4}$ & 0 & 76 & -- &15.5min \\ 
			\hline 
			\hline
			\multirow{2}{*}{Weak coupling}	& $10^{-3}$ & 0  & 12 & 163 & 1h20min  \\ 
			\cline{2-6} 
			& $10^{-4}$ & 30 & 122 & 994& 7h33min \\ 
			\hline 
			\hline
			\multirow{2}{*}{Full coupling}	& $10^{-3}$ & 0 & 13 & 344  & 1h53min \\ 
			\cline{2-6}
			& $10^{-4}$ & 0 & 381 & 5659 & 27h27min\\ 
			\hline 
		\end{tabular} 
	\end{center}
	\caption{Time integration in the whole cycle according to the coupling approach (ATS: additional time steps)}
	\label{tab:inteCycle_couplages}
\end{table}

\begin{figure}[ht]
	\subfloat[{Solution obtained by weak coupling.} \label{fig:mapP}]{
		\includegraphics[width=0.4\textwidth]{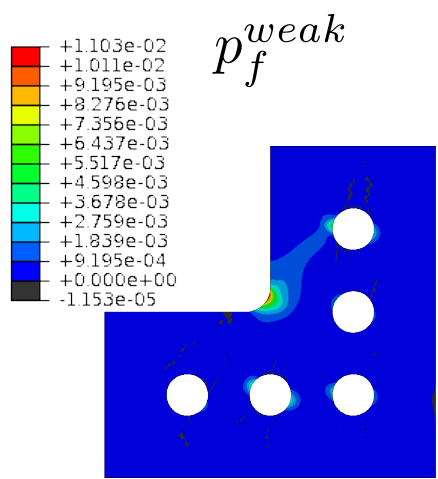}
	}
	\hfill
	\subfloat[Difference with monolithic solution.\label{fig:mapdiffP}]{
		\includegraphics[width=0.45\textwidth]{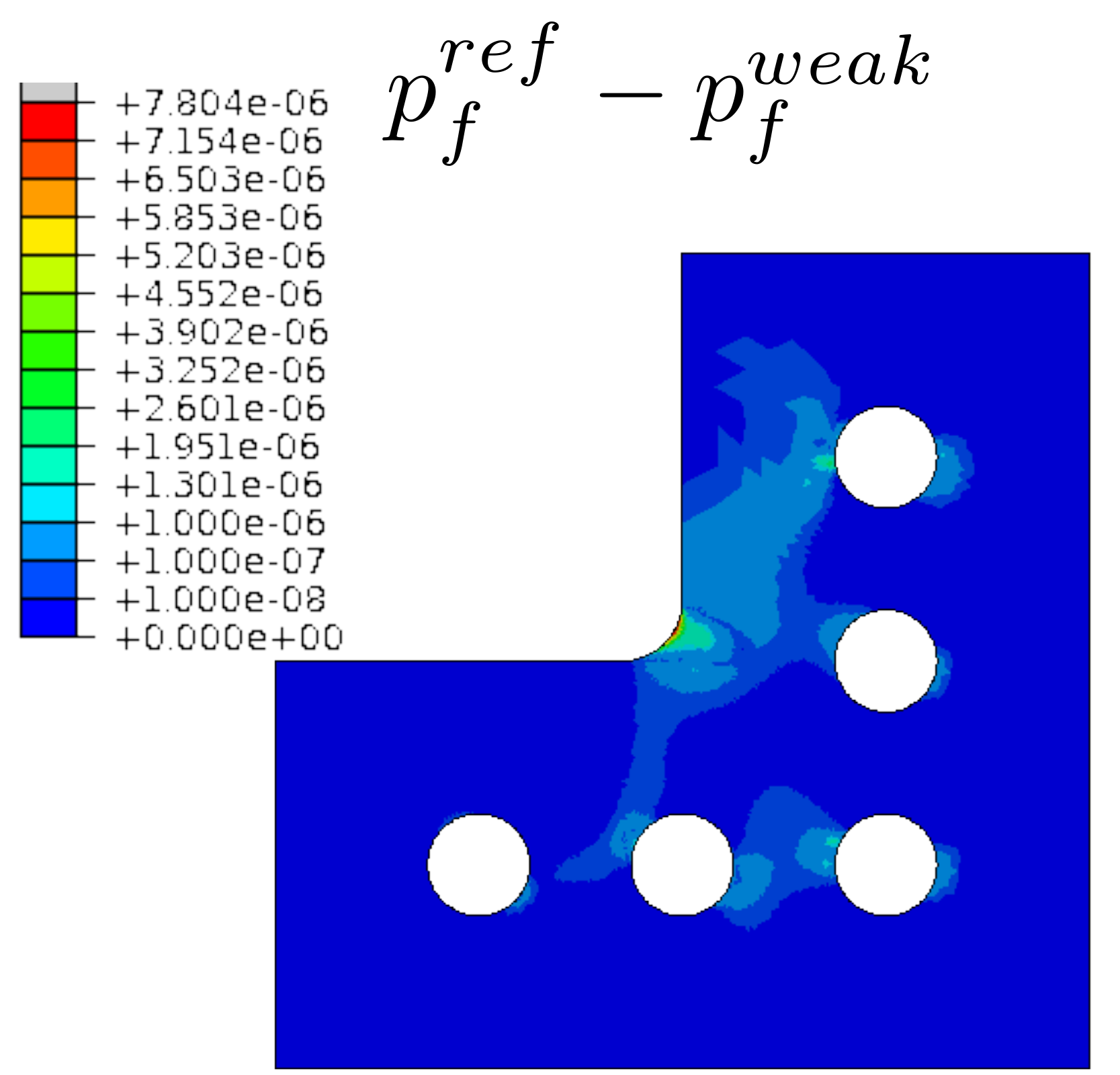}
	}
	\caption{{Weak coupling: distribution of fast plastic strain and comparison with monolithic solution in the zone of interest (${\Delta p_{\max}}=10^{-4}$)}.}
	\label{fig:errorMapP}
\end{figure}

Regarding the precision, according to the elementary tests of Subsection~\ref{sec:autotime} and those presented in Subsection~\ref{globaltimediscr},  the threshold  ${\Delta p_{\max}} = 10^{-4}$ is used. {The couplings are assessed with respect to the reference monolithic solution integrated with ${\Delta p_{\max}} = 10^{-5}$.} Table~\ref{tab:recapmaxLoad} summarizes the results obtained for the different coupling methods. 
Of course the submodeling approach is attractive in term of CPU but it underestimates the level of plastic strains by $38\%$, whereas the coupling strategies maintain an error level below $1\%$. The weak coupling leads to what is considered in practice as an acceptable level of accuracy compared to the overkill solution.  % (\textit{cf} Figure~\ref{fig:precCycleElem}).% As expected, it converges to the monolithic solution integrated with ${\Delta p_{\max}} = 10^{-4}${, with same maximum values obtained (as weak coupling) but computed in only 7 minutes}.
Figure~\ref{fig:errorMapP} shows that, as expected, largest differences are located near the corner and near the most loaded holes. 

\begin{table}[ht]
	\begin{center}
		\begin{tabular}{|c||c|c|c||c|}
			\hline 
			& {\textit{Submodeling}} & {Weak coupl.} & {Full coupl.}& Reference\\ 
			\hline \hline  
			& error $\left[\%\right]$   & error $ \left[\%\right]$ & error $ \left[\%\right]$ &$(\Delta p_{\max}=10^{-5})$\\ 
			\hline
			Mises	& 6.3 &  0.16 &  0.12 &  650.9 MPa\\ 
			\hline 
			$p_f$	& 38 &  0.8 &  0.08 & $1.05 \; 10^{-2}$\\ 
			\hline \hline
			Total time &{15min} & {7h33min} & {27h27min} & 1h11min \\
			\hline 
		\end{tabular} 
	\end{center}
	\caption{True error on the most loaded element at the end of the cycle and CPU time for various coupling strategies}
	\label{tab:recapmaxLoad}
\end{table}

%\begin{table}[ht]
%	\begin{center}
%		\begin{tabular}{|c||c|c|c|c|c|c|c|}
%			\hline 
%			Approaches	& \multicolumn{2}{c}{\textit{Submodeling}} & \multicolumn{2}{c}{Weak coupl.} & \multicolumn{2}{c}{Full coupl.}& Reference\\ 
%			\hline 
%			Fields 	&Values & $\eta \left[\%\right]$   & Values & $\eta \left[\%\right]$ & Values & $\eta \left[\%\right]$ &$\Delta p_{max}=10^{-5}$\\ 
%			\hline \hline 
%			Mises [MPa]	&610.2 & 6.3 & 649.9 & 0.16 & 650.1 & 0.12 &  650.9 \\ 
%			\hline 
%			$p_f$	&$6.54 \; 10^{-3}$& 38 & $1.06 \; 10^{-2}$ & 0.8 & $1.05 \; 10^{-2}$ & 0.08 & $1.05 \; 10^{-2}$\\ 
%			\hline 
%			Total time &\multicolumn{2}{c|}{15min} & \multicolumn{2}{c|}{7h30min} & \multicolumn{2}{c|}{27h} & 1h11min \\
%			\hline 
%		\end{tabular} 
%	\end{center}
%	\caption{Comparison of various couplings over one cycle. True error on the most loaded element and CPU time.}
%	\label{tab:recapmaxLoad}
%\end{table}

%
%\begin{figure}[ht]
%	\subfloat[Von Mises stress  \label{fig:precS}]{
%		\includegraphics[width=0.49\textwidth]{figure_8a}
%	}
%	\hfill
%	\subfloat[{Cumulated fast plastic strain}\label{fig:precP}]{
%		\includegraphics[width=0.49\textwidth]{figure_8b}
%	}
%	\caption{Evolution of the ``true error'' over a cycle for various couplings}
%	\label{fig:precCycleElem}
%\end{figure}

\color{black}

%{On Figure~\ref{fig:precCycleElem} an unexpected aspect regarding the accuracy of the full coupling appears. In fact, due to the addition of times steps during the unloading phase, the time discretization becomes finer than the one used for a monolithic computation integrated at ${\Delta p_{\max}} = 10^{-5}$. This means that the monolithic model which we considered as the reference for the measurement of ``true errors'' is not the more precise, and the green curves of Figure~\ref{fig:precCycleElem} can not be exploited. This problem of loss of reference in nonlinear computations is rather common, see~\cite{ALL10} for instance.}

\subsection{Improvement of performance}
\label{optimization}
In order to reduce the computational cost of the global/local analysis, the number of iterations performed on each time step may be reduced. As seen on Figure~\ref{fig:trueErrorP}, the accuracy is not improved after 5 iterations even if the equilibrium residual could still be decreased. The stagnation grossly corresponds to a relative residual of $10^{-3}$ which will be used (instead of $10^{-5}$) in the following computations. Furthermore, Aitken's $\delta^2$ is applied.  In this configuration, the total computational time is reduced by a factor 3 (2 hours and 44 minutes for the weak coupling strategy) without significantly altering the accuracy.

\begin{remark}[Cost issue]
Presently the performance of the method is strongly penalized by the fact that,  for each resolution (global or local) a new independent computation of Abaqus needs to be launched. This implies three additional stages compared to the monolithic solution (apart from the first time step): verification of input data, building of the geometry, assembly of the problem. Those stages are very costly.  Abaqus development team is considering evolutions to mitigate these costs. 
\end{remark}

\section{Application on a 3D example of industrial complexity}
\label{sec:3D}
\subsection{Model definition}
\begin{figure}[ht]
\centering
\includegraphics[width=0.9 \textwidth]{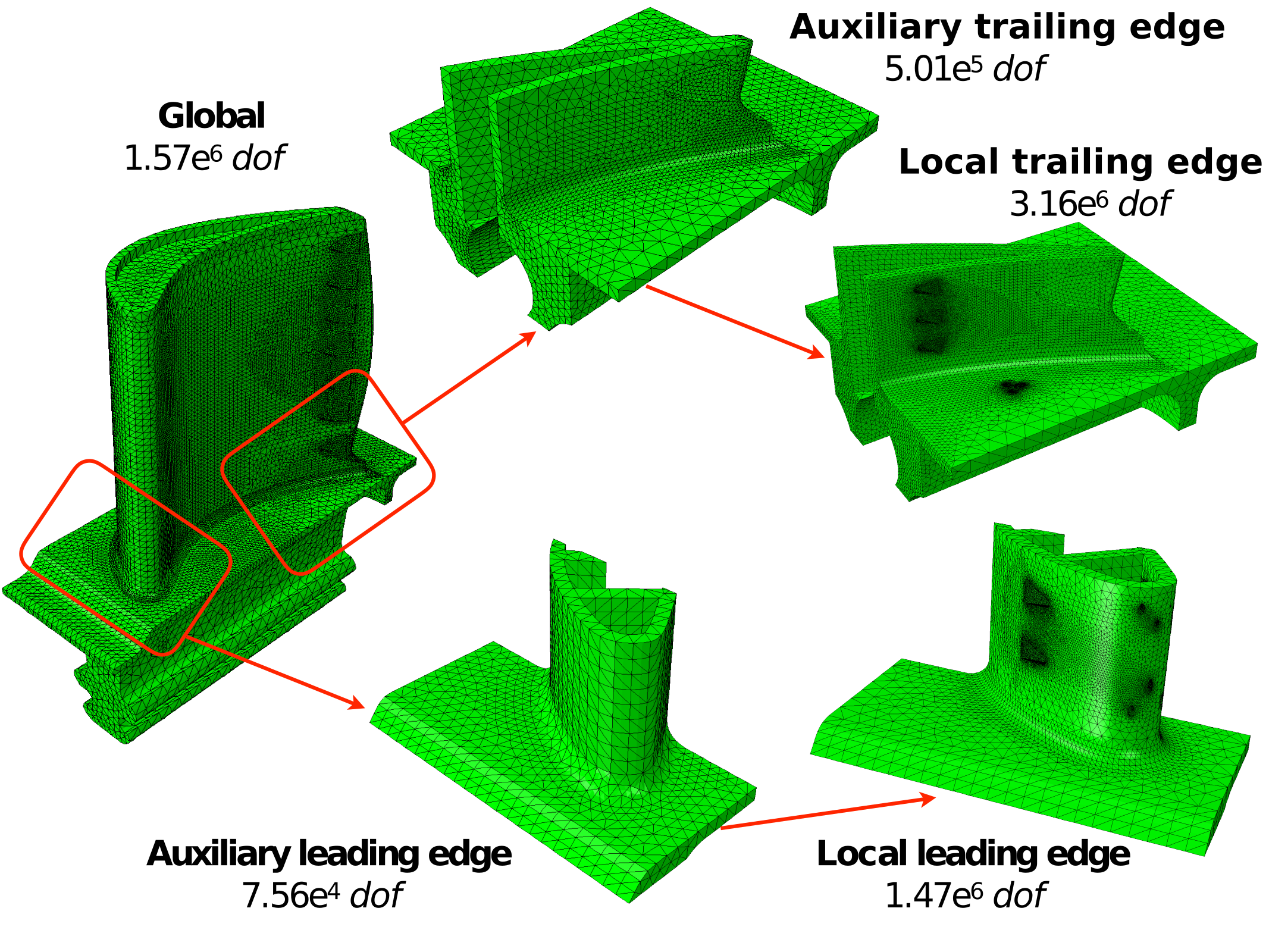}
\caption{Meshes related to 3D test case}
\label{3Dmeshes}
\end{figure}

The method is now applied on a 3D model representing a high pressure turbine blade close to actual industrial problems.  Figure~\ref{3Dmeshes} presents the meshes and gives the associated number of degrees of freedom.
 
Two local areas are introduced: one on the leading edge (LE) and another on the trailing edge (EE). In these two zooms, micro-perforations are added; they require meshes about ten times finer than the global one. Because the models remain of reasonable size ($5.63\, 10^6$ degrees of freedom), a monolithic approach can be computed, with 15 processors and 70Gb of memory, and be used as a reference.

\begin{figure}[ht]
	\subfloat[Leading edge interface\label{fig:interfaceBA}]{
		\includegraphics[width=0.4\textwidth]{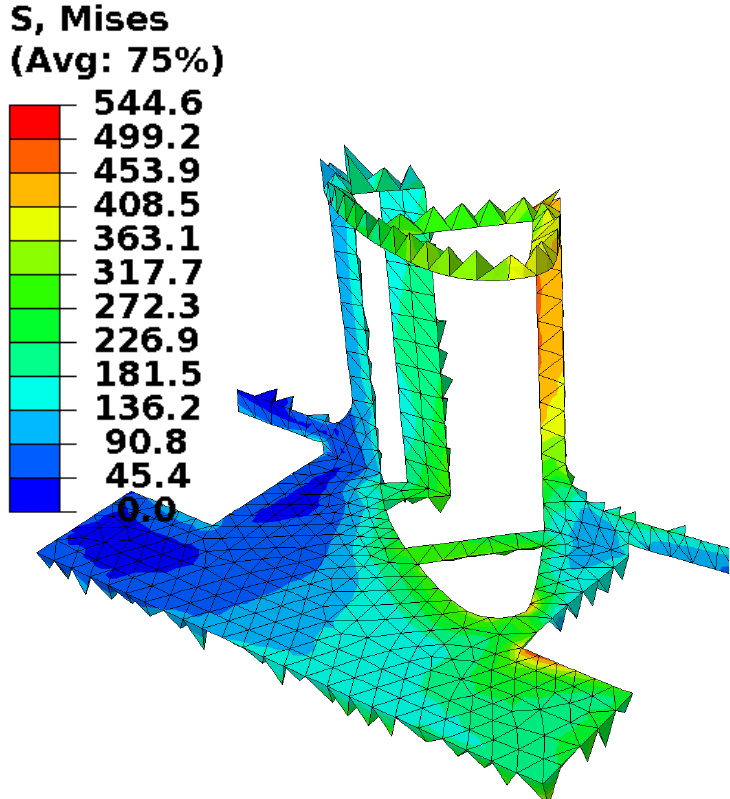}
	}
	\hfill
	\subfloat[Trailing edge interface \label{fig:interfaceBF}]{
		\includegraphics[width=0.4\textwidth]{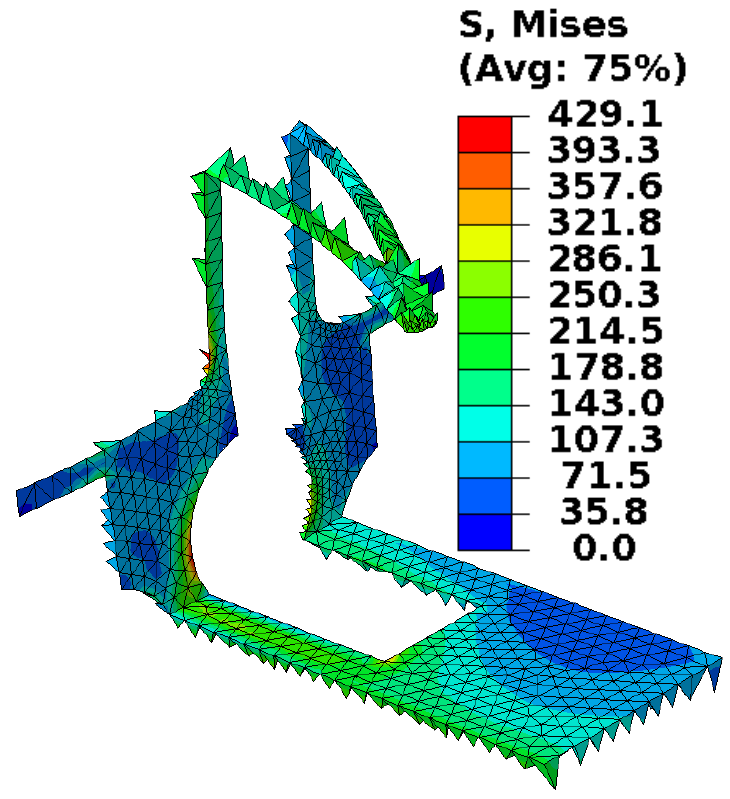}
	}
\caption{Interfaces between 3D models}
\label{fig:3D_Interfaces}
\end{figure}

In 3D, the relative size of the interfaces, which characterizes the cost of exchanges, is quite small. On this example they correspond only to 2580 and 2520 nodes (\textit{cf.} Figure~\ref{fig:3D_Interfaces}) and the cost of the exchanges at the interfaces becomes very cheap compared to the cost of the resolution of the different models. The local and complementary areas present non matching meshes, so the method presented in Section~\ref{ssec:non-matching} is applied. The reference model itself makes use the \texttt{TIE} coupling of \textit{Abaqus}. 

The material model, external loads and boundary conditions are similar to the 2D example: the blade is subjected to a pressure on the leading edge, centrifugal forces on all models and the normal component of the blade foot displacement is set to zero.
 
\subsection{Numerical results}
\begin{figure}[ht]
	\subfloat[Von Mises stress \label{fig:Glob-S}]{
		\includegraphics[width=0.4\textwidth]{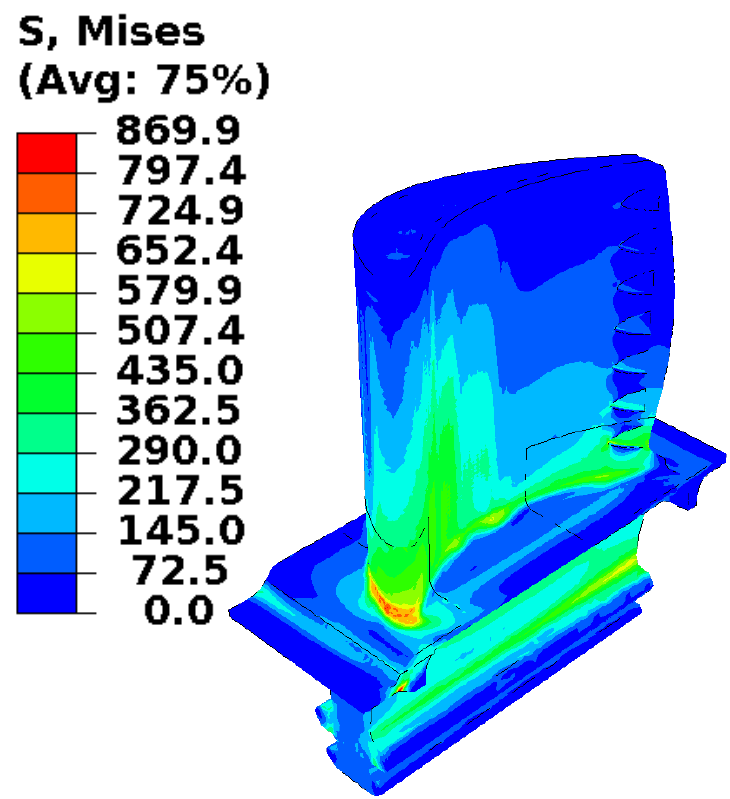}
	}
	\hfill
	\subfloat[{Cumulated fast plastic strain} \label{fig:Glob-P}]{
		\includegraphics[width=0.4\textwidth]{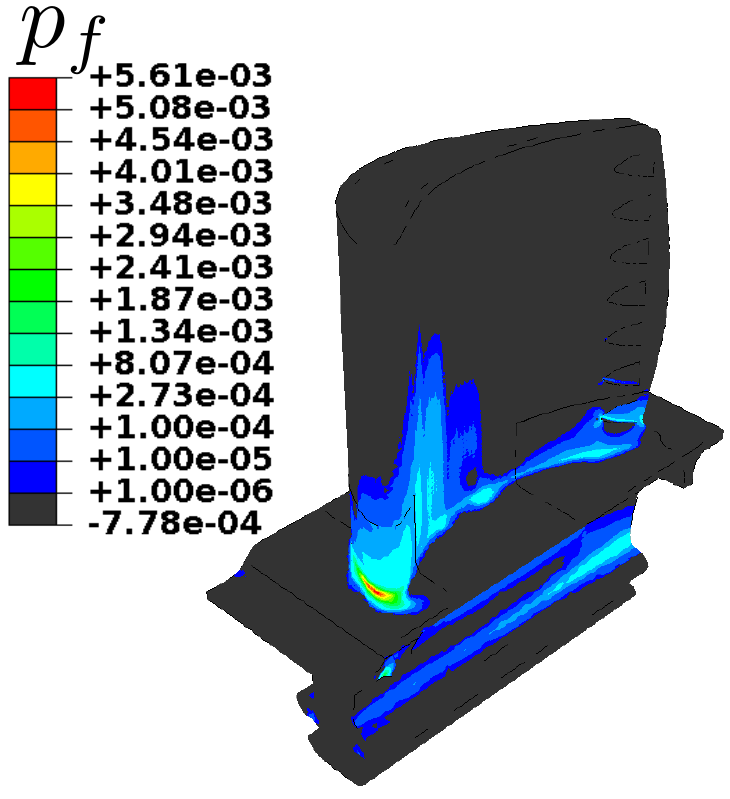}
	}
	\caption{Global solutions computed in 18 minutes}
	\label{fig:global}
\end{figure}

\begin{figure}[ht]
	\subfloat[Time integration performed \label{fig:timeInte3D}]{
		\includegraphics[width=0.49\textwidth]{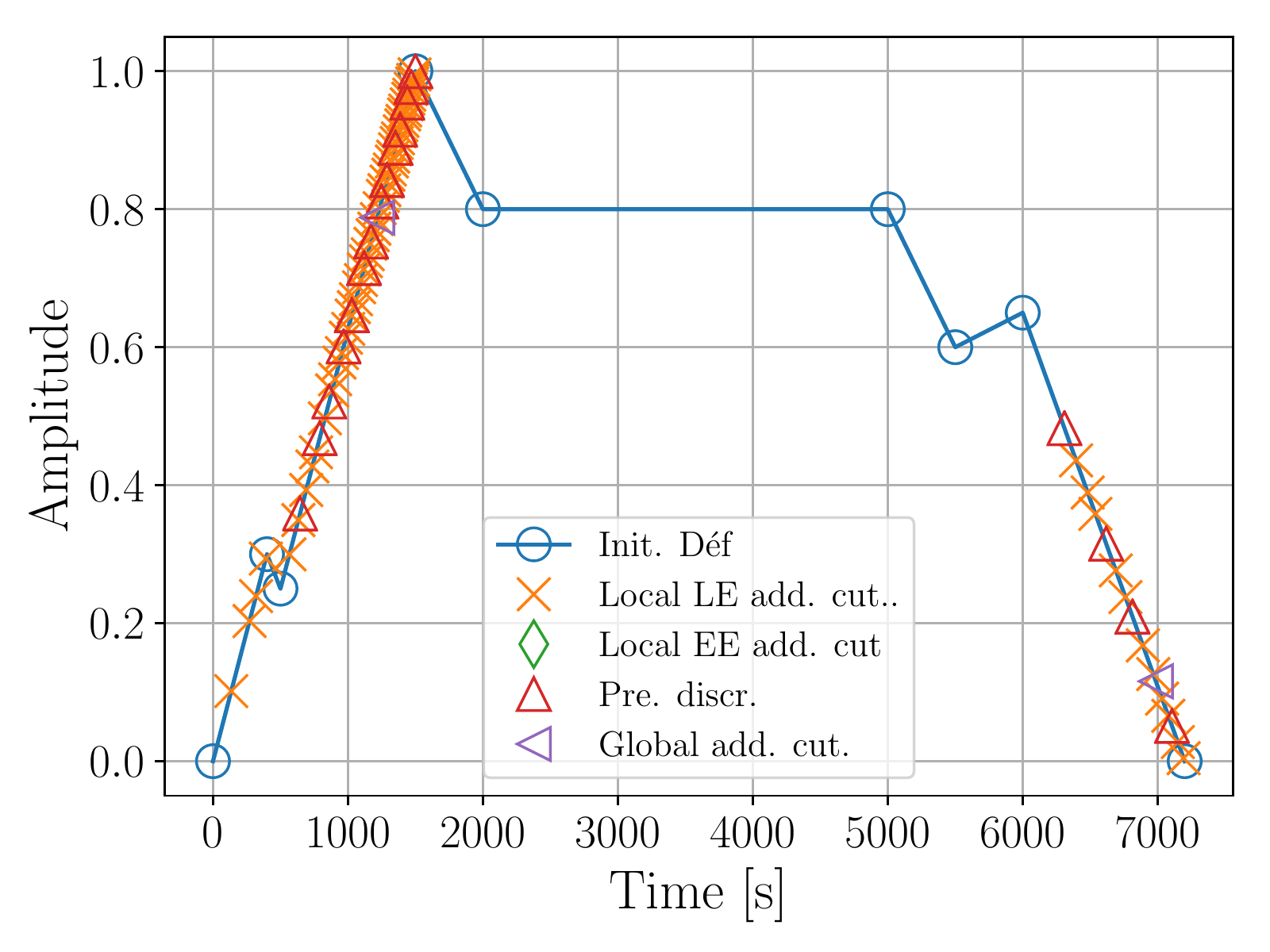}
	}
	\hfill
	\subfloat[Residual over coupling iteration (maximum load) \label{fig:residu3D}]{
		\includegraphics[width=0.49\textwidth]{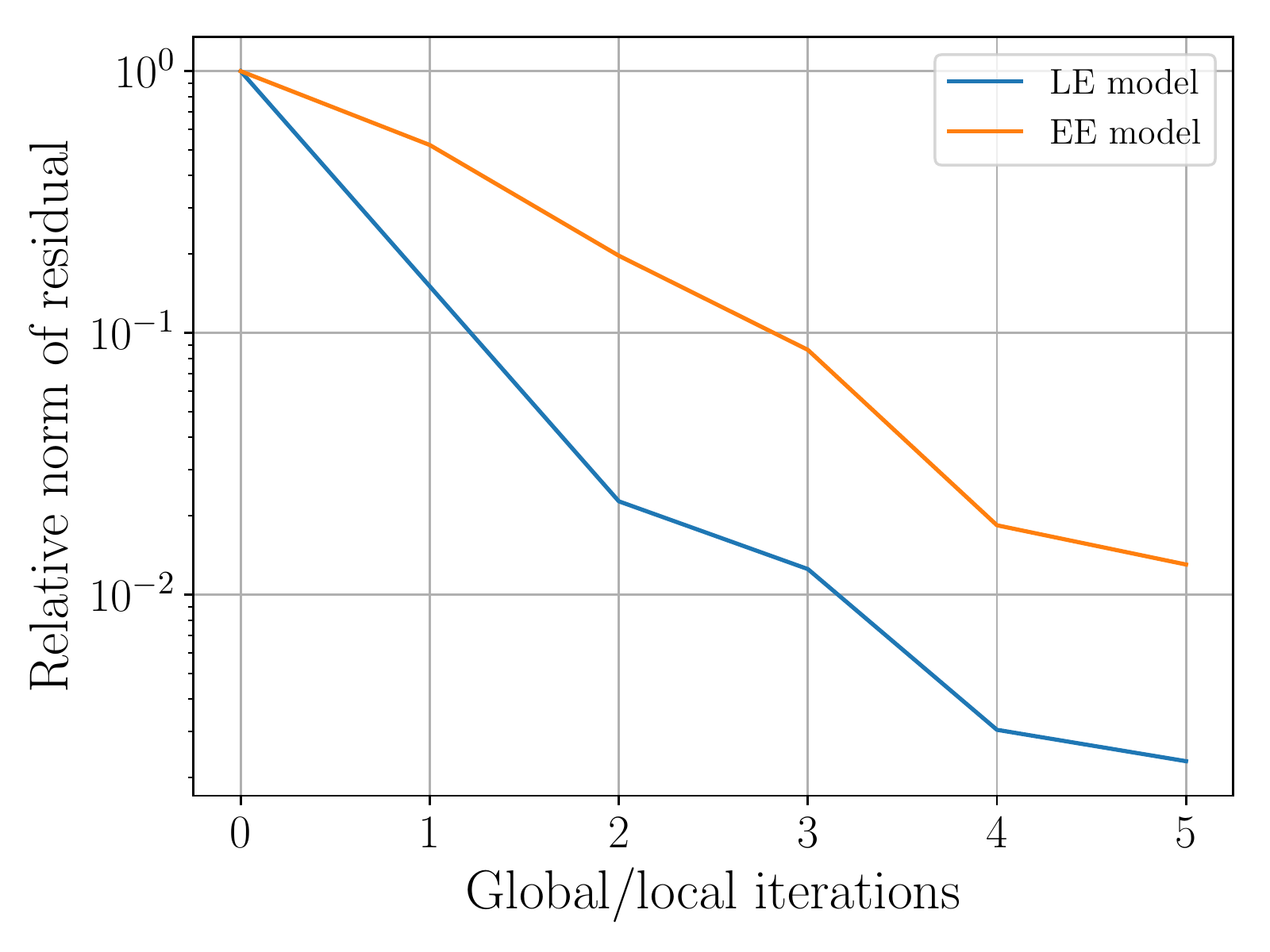}
	}
	\caption{Performance on 3D test case}
	\label{fig:timeIntegration}
\end{figure}

In order to evaluate the performance of the algorithm, the optimized settings are used and the number of coupling iterations is limited to 5.  Figure~\ref{fig:timeIntegration} displays the time steps added at the local and global levels for the threshold of ${\Delta p_{\max}} = 5\, 10^{-4}$. The local model of the leading edge requires 62 additional times steps whereas the local model on the trailing edge does not demand any time refinement. A large plastic area, which connects all local models is obtained, see Figure~\ref{fig:global}.

The convergence rate plotted on Figure~\ref{fig:residu3D} is comparable to the one obtained on 2D test case. %In this 3D test case, the complexity of the interface, which may slow down the convergence rate, is balanced by the smaller redistribution compared to the 2D simulation.
{Indeed the problem might appear more complex from a domain decomposition point of view because of the complex shape of the interface, but this difficulty is largely compensated by the lesser	 redistribution of nonlinearity compared to the 2D case.}

\subsubsection{Accuracy aspects}
\begin{figure}[ht]
	\subfloat[Error in von Mises stress \label{fig:error3DSig}]{
		\includegraphics[width=0.49\textwidth]{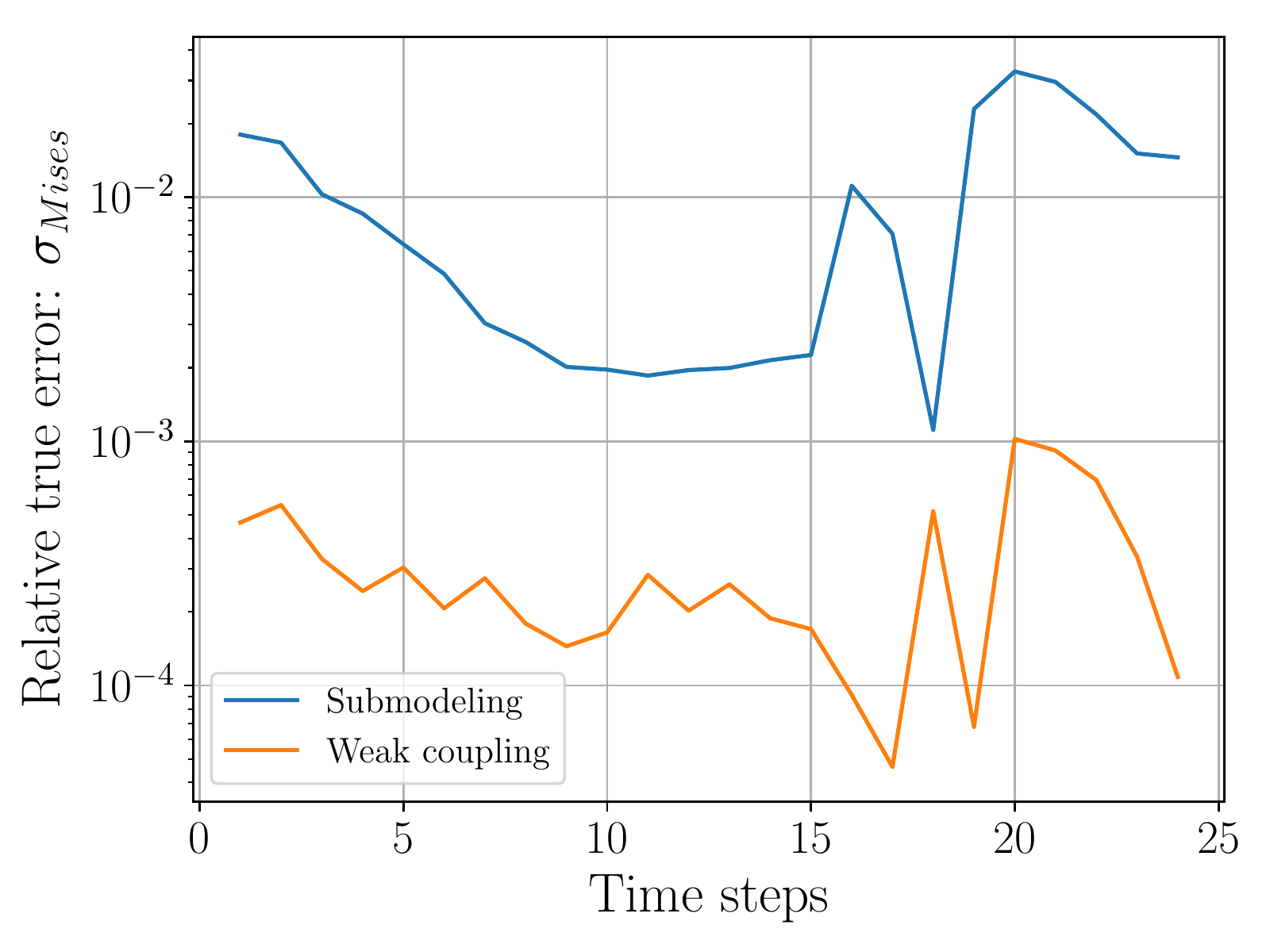}
	}
	\hfill
	\subfloat[{Error in cumulated fast plastic strain}\label{fig:error3DPlast}]{
		\includegraphics[width=0.49\textwidth]{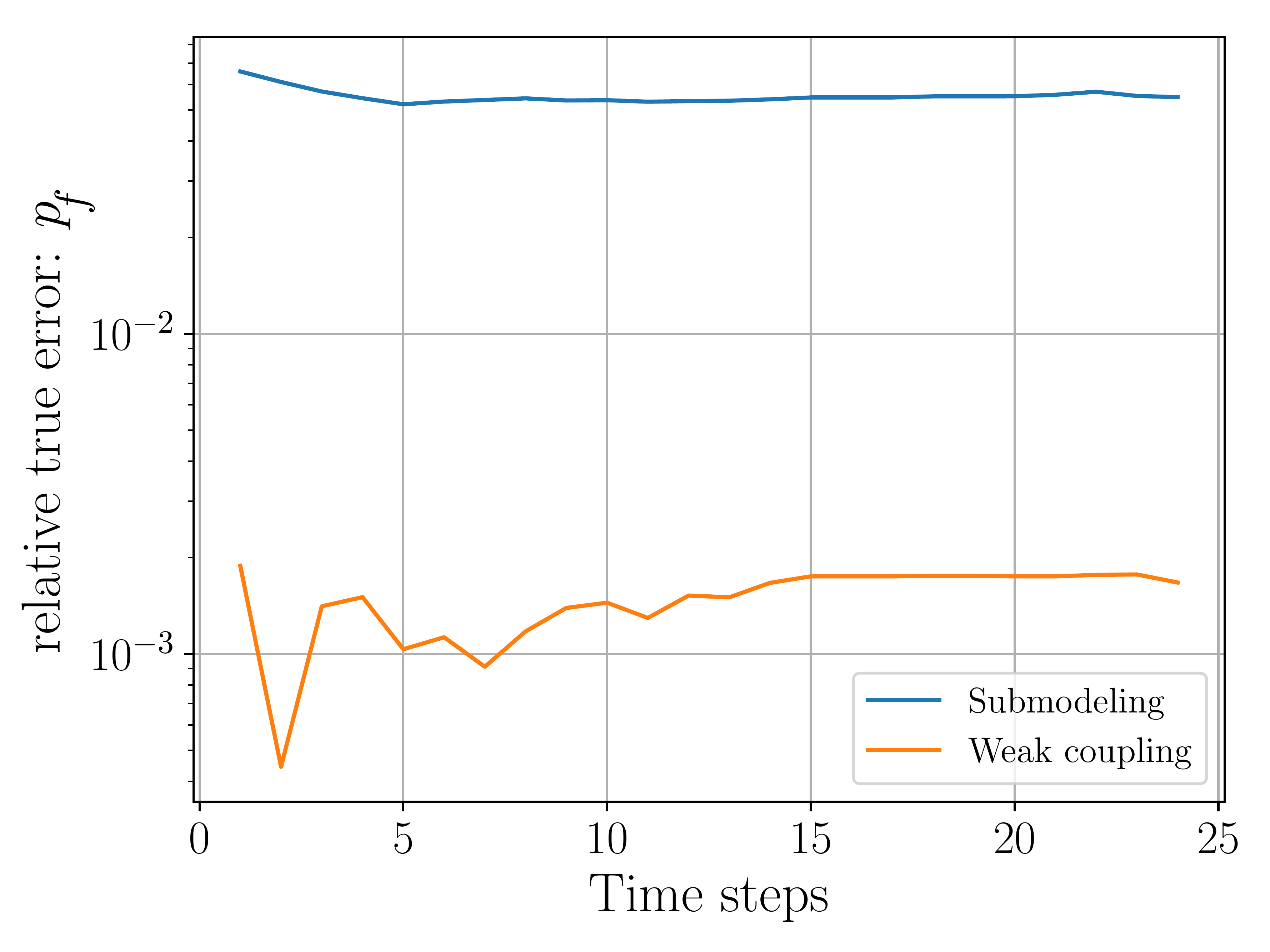}
	}
	\caption{Error level over the cycle on the most loaded element of the leading edge}
	\label{fig:errorLevel}
\end{figure}

A satisfying level of accuracy is obtained everywhere. Figure~\ref{fig:errorLevel} reports error levels for the most loaded element. Once again the error of the submodeling approach is high, although smaller than in the 2D case. This is because the global to local redistributions are less important in this case.

\subsubsection{Performance}
The reference (monolithic) model is computed in 7 hours and 27 minutes and the submodeling provides a solution on the cycle in only 4 hours and 37 minutes.  The weak coupling applied in its optimal configuration solves the model in 31 hours and 15 minutes, representing a ratio of 4 with the reference. As a reminder the ratio was about 24 on the 2D test case. But, as already discussed in Subsection~\ref{optimization}, a fair comparison will be possible  only when the issue of the model reconstruction at each time step will be solved.

\section*{Conclusion}
\label{conclusion}
The global/local non invasive method has been extended to allow a proper integration of the constitutive relation on each domain to be coupled. The use of different time grids over the iterations and depending on the domains seems mandatory. Two strategies have been compared and in practice the lighter one, called weak coupling, seems to offer a good compromise between cost and precision. Still, in a 3D case where a monolithic reference could be computed, the coupling strategy was 4 times slower. Let us recall here that one of the motivation is to allow the engineer to easily modify some local details without having to reconstruct the whole mesh. In fact the later operation is really cumbersome for complex industrial parts and is more and more often externalized. Moreover it will be possible to properly assess the performance of the method on Abaqus only when it will allow to construct the different models only once, a possibility offered by codes like Code\_Aster for example.
Departing from the context of this paper, it is also clear that using some guaranteed error estimation techniques would be a real asset to the method. It would allow to adapt the choice of the integration threshold ${\Delta p_{\max}}$ and the choice of the convergence criterion for  the global/local iterations, for a given target accuracy.  A huge literature devoted to error estimation is of interest here \cite{cha-Lad2009,Lad-cha2010}, in particular regarding quantities of interest,  but it should be ported to the context of non-invasive tools. 

\section*{Acknowledgement}
{\color{black}
The authors are extremely grateful for the helpful and benevolent remarks given by the reviewers. Moreover the expertise of Reviewer~3 on Abaqus appeared to be a real asset.
}

\bibliography{biblio-coupling.bib}

\begin{thebibliography}{10}

\bibitem{allix_2001}
O.~Allix.
\newblock A composite damage meso-model for impact problems.
\newblock {\em Composites Science and Technology}, 61:2193--2205, 2001.

\bibitem{ZSE15}
Armines and Onera.
\newblock {\em Z-set, Material and Structure Analysis Suite, Materials manual
  Version 8.6}, 2015.

\bibitem{BET14-1}
Omar Bettinotti, Olivier Allix, and Beno\^{i}t Malherbe.
\newblock A coupling strategy for adaptive local refinement in space and time
  with a fixed global model in explicit dynamics.
\newblock {\em Computational Mechanics}, 53(4):561--574, 2014.

\bibitem{BET17}
Omar Bettinotti, Olivier Allix, Umberto Perego, Victor Oancea, and Beno\^{i}t
  Malherbe.
\newblock Simulation of delamination under impact using a global local method
  in explicit dynamics.
\newblock {\em Finite Elements in Analysis and Design}, 125(8):1--13, 2017.

\bibitem{BLA17-2}
Maxime Blanchard.
\newblock {\em M{\'e}thode global/local non-intrusive pour les simulations
  cycliques non-lin{\'e}aires}.
\newblock PhD thesis, {\'E}cole Normale Sup{\'e}rieure Paris-Saclay, 2017.

\bibitem{CHA89}
Jean-Louis Chaboche.
\newblock Constitutive equations for cyclic plasticity and cyclic
  viscoplasticity.
\newblock {\em International journal of plasticity}, 5(3):247--302, 1989.

\bibitem{cha-Lad2009}
L.~Chamoin and P.~Ladev{\`e}ze.
\newblock Strict and practical bounds through a non-intrusive and goal-oriented
  error estimation method for linear viscoelasticity problems.
\newblock {\em Finite Element Analysis and Design}, 45:251--262, 2009.

\bibitem{COR99}
N.~G. Cormier, B.~S. Smallwood, G.~B. Sinclair, and G.~Meda.
\newblock Aggressive submodelling of stress concentrations.
\newblock {\em International Journal for Numerical Methods in Engineering},
  46(6):889--909, 1999.

\bibitem{DHI98}
Hachmi~Ben Dhia.
\newblock Probl\`{e}mes m\'{e}caniques multi-\'{e}chelles: la m\'{e}thode
  {A}rlequin.
\newblock {\em Comptes Rendus de l'Acad\'{e}mie des Sciences - Series IIB -
  Mechanics-Physics-Astronomy}, 326(12):899 -- 904, 1998.

\bibitem{DUV18}
M.~Duval, A.~Lozinski, J.~C. Passieux, and et~al.
\newblock Residual error based adaptive mesh refinement with the non-intrusive
  patch algorithm.
\newblock {\em Computer Methods in Applied Mechanics and Engineering},
  329:118--143, 2018.

\bibitem{DUV16}
Micka{\"e}l Duval, Jean-Charles Passieux, Michel Sala{\"u}n, and St{\'e}phane
  Guinard.
\newblock Non-intrusive coupling: recent advances and scalable nonlinear domain
  decomposition.
\newblock {\em Archives of Computational Methods in Engineering}, 23(1):17--38,
  2016.

\bibitem{FAR91}
Charbel Farhat and Francois-Xavier Roux.
\newblock A method of finite element tearing and interconnecting and its
  parallel solution algorithm.
\newblock {\em International Journal for Numerical Methods in Engineering},
  32(6):1205--1227, 1991.

\bibitem{FIL18}
Treavis~B. Fillmore and C.~Armando Duarte.
\newblock A hierarchical non-intrusive algorithm for the generalized finite
  element method.
\newblock {\em Advanced Modeling and Simulation in Engineering Sciences}, 5(2),
  2018.

\bibitem{GEN11}
L~Gendre, O~Allix, and P~Gosselet.
\newblock A two-scale approximation of the {S}chur complement and its use for
  non-intrusive coupling.
\newblock {\em International Journal for Numerical Methods in Engineering},
  87(9):889--905, 2011.

\bibitem{GEN09a}
Lionel Gendre, Olivier Allix, Pierre Gosselet, and Fran{\c{c}}ois Comte.
\newblock Non-intrusive and exact global/local techniques for structural
  problems with local plasticity.
\newblock {\em Computational Mechanics}, 44(2):233--245, 2009.

\bibitem{GLO05}
Roland Glowinski, Jiwen He, Alexei Lozinski, Jacques Rappaz, and Jo{\"e}l
  Wagner.
\newblock Finite element approximation of multi-scale elliptic problems using
  patches of elements.
\newblock {\em Numerische Mathematik}, 101(4):663--687, 2005.

\bibitem{GOS17}
Pierre Gosselet, Maxime Blanchard, and Olivier Allix.
\newblock Non-invasive global-local coupling as a schwarz domain decomposition
  method: acceleration and generalization.
\newblock {\em Advanced Modeling and Simulation in Engineering Sciences}, 5(4),
  2018.
\newblock hal-01613966v1.

\bibitem{GUG14}
Guillaume Guguin, Olivier Allix, Pierre Gosselet, and St{\'e}phane Guinard.
\newblock Nonintrusive coupling of 3d and 2d laminated composite models based
  on finite element 3d recovery.
\newblock {\em International Journal for Numerical Methods in Engineering},
  98(5):324--343, 2014.

\bibitem{GUI18}
St\'ephane Guinard, Robin Bouclier, Mateus Toniolli, and Jean-Charles Passieux.
\newblock Multiscale analysis of complex aeronautical structures using robust
  non-intrusive coupling.
\newblock {\em Advanced Modeling and Simulation in Engineering Sciences}, 5(1),
  2018.

\bibitem{HOL13}
M.~Holl, S.~Loehnert, and P~Wriggers.
\newblock An adaptive multiscale method for crack propagation and crack
  coalescence.
\newblock {\em International Journal for Numerical Methods in Engineering},
  93(1):23--51, 2013.

\bibitem{HUE16}
S~H\"{u}hne, J~Reinoso, Eelco Jansen, and et~al.
\newblock A two-way loose coupling procedure for investigating the buckling and
  damage behaviour of stiffened composite panels.
\newblock {\em Composite Structures}, 136:513--525, 2016.

\bibitem{LAD99}
P.~Ladev{\`e}ze.
\newblock {\em Nonlinear Computanional Structural Mechanics. New approaches and
  Non-Incremental Methods of calculation}.
\newblock Springer-Verlag, 1999.

\bibitem{Lad-cha2010}
P.~Ladev{\`e}ze, L.~Chamoin, and E.~Florentin.
\newblock A new non-intrusive technique for the construction of admissible
  stress fields in model verification.
\newblock {\em Computer Methods in Applied Mechanics and Engineering},
  199:766--777, 2010.

\bibitem{LEM09}
J.~Lemaitre, J.L. Chaboche, A.~Benallal, and R.~Desmorat.
\newblock {\em M{\'e}canique des mat{\'e}riaux solides
  3$\,^{\circ}${\'e}dition}.
\newblock Dunod, 2009.

\bibitem{LIU14}
Y.J. Liu, Q.~Sun, and X.L. Fan.
\newblock A non-intrusive global/local algorithm with non-matching interface:
  Derivation and numerical validation.
\newblock {\em Comput. Methods Appl. Mech. Engrg.}, 277:81--103, 2014.

\bibitem{LON13}
A.~Longuet, A.~Burteau, A.~Comte, and A.~Crouchez-Pilot.
\newblock Incremental lifing method applied to high temperature aeronautical
  component.
\newblock In {\em Actes du 11eme colloque national en calcul des structures},
  pages 703--708, 2013.

\bibitem{MAN93}
Jan Mandel.
\newblock Balancing domain decomposition.
\newblock {\em Communications in numerical methods in engineering},
  9(3):233--241, 1993.

\bibitem{NOU17}
Anthony Nouy and Florent Pled.
\newblock A multiscale method for semi-linear elliptic equations with localized
  uncertainties and non-linearities.
\newblock {\em ESAIM: Mathematical Modelling and Numerical Analysis}, 2018.

\bibitem{PAS13}
Jean-Charles Passieux, Julien R{\'e}thor{\'e}, Anthony Gravouil, and
  Marie-Christine Baietto.
\newblock Local/global non-intrusive crack propagation simulation using a
  multigrid x-fem solver.
\newblock {\em Computational Mechanics}, 52(6):1381--1393, 2013.

\bibitem{PLE12}
J.~Plews, C.~A. Duarte, and T.~Eason.
\newblock An improved non-intrusive global-local approach for sharp thermal
  gradients in a standard fea platform.
\newblock {\em International Journal for Numerical Methods in Engineering},
  91(4):361--397, 2012.

\bibitem{RUE14}
Martin Ruess, Dominik Schillinger, Ali~I {\"O}zcan, and Ernst Rank.
\newblock Weak coupling for isogeometric analysis of non-matching and trimmed
  multi-patch geometries.
\newblock {\em Computer Methods in Applied Mechanics and Engineering},
  269:46--71, 2014.

\bibitem{ABA16}
Abaqus Simulia.
\newblock Abaqus documentation: Analysis user's guide.
\newblock {\em Providence, RI USA}, 2016.

\bibitem{TEM11}
I.~Temizer and P.~Wriggers.
\newblock An adaptive multiscale resolution strategy for the finite deformation
  analysis of microheterogeneous structures.
\newblock {\em Computer Methods in Applied Mechanics and Engineering},
  200(37-40):2639 -- 2661, 2011.
\newblock Special Issue on Modeling Error Estimation and Adaptive Modeling.

\end{thebibliography}
\end{document}